\documentclass{aa}  

\usepackage{graphicx}
\usepackage{natbib}
\bibpunct{(}{)}{;}{a}{}{,}
\newcommand{\tex}{\ensuremath{T_{\rm ex}}}

\usepackage{txfonts}
\begin{document}

   \title{The origin of gas-phase HCO and CH$_3$O radicals in prestellar cores\thanks{Based on observations carried out with the IRAM 30m telescope. IRAM is supported by
INSU/CNRS (France), MPG (Germany), and IGN (Spain)}}

   \author{A. Bacmann
          \inst{1,2}
          \and
          A. Faure\inst{1,2}
          }

   \institute{Univ. Grenoble Alpes, IPAG, F-38000 Grenoble, France 
         \and
             CNRS, IPAG, F-38000 Grenoble, France\\
             \email{aurore.bacmann@ujf-grenoble.fr, alexandre.faure@ujf-grenoble.fr}
             }

   \date{Received September 15, 1996; accepted March 16, 1997}

 
  \abstract
   {The recent unexpected detection of terrestrial complex organic molecules in the cold ($\sim$\,10\,K) gas has cast doubts on the 
   commonly accepted formation mechanisms of these species. Standard gas-phase mechanisms are inefficient and tend to underproduce these molecules, and many of the key reactions involved are unconstrained. Grain-surface mechanisms, which were presented as a viable alternative, suffer from the fact that they rely on grain surface diffusion of heavy radicals, which is not possible thermally at very low temperatures.} 
   {One of the simplest terrestrial complex organic molecules,
methanol is believed to form on cold grain surfaces following from successive H atom additions on CO. Unlike heavier species, H atoms are very mobile on grain surfaces even at 10\,K. Intermediate species involved in grain surface methanol formation by CO hydrogenation are the radicals HCO and CH$_3$O, as well as the stable species formaldehyde H$_2$CO. These radicals are thought to be precursors of complex organic molecules on grain surfaces. }
   {We present new observations of the HCO and CH$_3$O radicals in a sample of prestellar cores and carry out an analysis of the abundances of the species HCO, H$_2$CO, CH$_3$O, and CH$_3$OH, which represent the various stages of grain-surface hydrogenation of CO to CH$_3$OH.}
{The abundance ratios between the various intermediate species in the
  hydrogenation reaction of CO on grains are similar in all sources of
  our sample, HCO\,:\,H$_2$CO\,:\,CH$_3$O\,:\,CH$_3$OH $\sim
  10:100:1:100$. We argue that these ratios may not be representative of   the primordial abundances on the grains but, rather, suggest that
  the radicals HCO and CH$_3$O are gas-phase products of the
  precursors H$_2$CO and CH$_3$OH, respectively. Various gas-phase
  pathways are considered, including neutral-neutral and ion-molecule
  reactions, and simple estimates of HCO and CH$_3$O abundances
  are compared to the observations. Critical reaction rate constants,
  branching ratios, and intermediate species are finally identified.}
   {}

\titlerunning{HCO and CH$_3$O in prestellar cores}


   \maketitle
%

\section{Introduction}

Complex organic molecules (hereafter COMs) have long been detected in the interstellar medium \citep{Gottlieb:1973p5025,Brown:1975p3219,Tucker:1974fs,Avery:1976eq}  and are known to present a wide variety of species from insaturated carbon chains like C$_{4}$H or cyanopolyynes to (more) saturated, terrestrial-like O- or N- bearing molecules, such as methyl formate (CH$_3$OCHO) or acetonitrile (CH$_3$CN). If carbon chain molecules were readily found in dark clouds, and their abundances reasonably accounted for by cold gas-phase chemistry \cite[see][for a recent review]{Loison:2014p4321}, terrestrial COMs have until recently been thought to be characteristic of hot cores or hot corinos, where they were found to be abundant using both single-dish and interferometric observations \citep{vanDishoeck:1995p3020,Cazaux:2003p2499,Bottinelli:2004p1209}. 

For these COMs, the most favoured formation mechanism involve the freezing out of simple atoms and simple molecules like CO onto the dust grains during the prestellar phase, where they can be hydrogenated. After the protostar forms and starts heating up its surroundings, the radicals on the grain mantles, which were either created by UV irradiation of the ices \citep{Garrod:2006p2498} or trapped in grain mantles \citep{Taquet:2012p5062}, start diffusing on the surface at a temperature of about 30\,K. Upon reaction, these radicals can form more complex molecules like methyl formate or dimethyl ether. The synthesised molecules are then thermally evaporated at higher temperatures. 

Pure gas-phase formation mechanisms involving the reactions of protonated methanol or protonated formaldehyde with neutrals followed by dissociative recombinations \citep{Charnley:1992p3105} have been shown to lead to COM abundances that are several orders of magnitude lower than those observed, because the formation of complex molecular ions is not energetically favourable \citep{Horn:2004p2926},  and dissociative recombinations favour the channels disrupting the molecular ions into small fragments \citep[see][in the case of methanol]{Geppert:2006p4196}. The  detections of saturated COMs in a prestellar core at 10\,K by \citet{Bacmann:2012p4015} or in the cold source B1-b by \citet{Cernicharo:2012p4397} 
have, however, cast doubt on grain surface mechanisms, since the temperature is not high enough to provide radicals with the necessary amount of energy to diffuse on the grains. 

A possible source of energy within prestellar cores could be cosmic rays, either directly or indirectly with secondary UV photons, but these remain widely unconstrained, and it is also unclear whether such mechanisms would produce COM amounts consistent with those observed in the gas-phase. Recently, \citet{Vasyunin:2013p4398} and \citet{Balucani:2015p5368} have proposed new gas-phase mechanisms based in particular on radiative association reactions, which are able to account for several COM abundances in prestellar cores (though not for e.g. methyl formate). Alternatively, \citet{Ruaud:2015p5367} studied the possibility that COMs form on grain surfaces following the direct chemical reaction of carbon atoms with the ice mantle constituents. However promising, those models rely on poorly constrained physical mechanisms  \citep[e.g. chemical desorption:][]{Garrod:2007p2821} and reactions where the rates are very uncertain and should be measured in the laboratory for confirmation. The question of the formation mechanism of COMs at very cold temperatures remains debated and unsettled.  

Methanol is one of the simplest terrestrial COMs seen in the interstellar medium. It is widely detected in cold dark clouds and prestellar cores \citep{Friberg:1988p3131,Tafalla:2006p3988}, and its formation mechanism is better constrained than that of larger COMs because more experimental data are available. Gas-phase chemistry is believed to be inefficient at forming CH$_3$OH \citep{Garrod:2006p4851,Geppert:2006p4196} and to underestimate observed methanol abundances of 10$^{-9}$ by several orders of magnitude. These estimates rely partly on the value of the branching ratios of the dissociative recombination of protonated methanol, which was measured experimentally by \citet{Geppert:2006p4196} and yields only 3\% for $\mathrm{CH_3OH_2^+ + e^- \rightarrow CH_3OH}$. Grain-surface reactions, on the other hand, provide an alternative way to form methanol in the cold interstellar medium, since CO  can be successively hydrogenated on the grains following $\mathrm{CO \rightarrow HCO \rightarrow H_2CO \rightarrow CH_3O \rightarrow CH_3OH}$ \citep[e.g.][]{Brown:1988p3318}. 

Laboratory experiments carried out at 10\,K have confirmed the possibility of synthesising methanol on grain surfaces by injecting H atoms towards a cold CO surface \citep{Watanabe:2002p3383}.  According to \citet{Watanabe:2002p3383} this mechanism is efficient, and as much as 10\% of the CO could be converted to CH$_3$OH. The products of this reaction are assumed to be desorbed back to the gas phase by cosmic ray impact or irradiation by secondary UV photons in regions where the temperature is not high enough to warrant evaporation of methanol ices (at $\sim 100$\,K). The process of methanol formation by multiple hydrogenation of CO is most efficient in very cold gas (10-15\,K) because higher temperatures mean that H atoms will not remain on grain surfaces, and CO molecules will also return to the gas phase above $20-30$\,K. Therefore, it is presently thought that the methanol seen in the hot cores of young low-mass stars has its origin in the prestellar stage.

Intermediate products in this process like HCO and CH$_3$O are of particular interest, since they are believed to be the precursors of saturated COMs in the kind of scenario proposed by \citet{Garrod:2006p2498}, for example. Though the formyl radical HCO was detected nearly 40 years ago \citep{Snyder:1976p4306}, observations  in the cold gas are still scarce. Mostly, HCO has been detected in moderate density or photon-dominated regions. \citet{Gerin:2009p4072} observed HCO in the PDR  of the Horsehead and also at one position in the dense core adjacent to the PDR. \citet{Frau:2012p3985} also observed a series of cold starless cores in the Pipe nebula. HCO is seen towards the  cores with modest column densities ($N\sim10^{22}$\,cm$^{-2}$) but not in the denser (probably prestellar) deuterated cores. The methoxy radical CH$_3$O was only detected recently towards the cold cloud B1-b \citep{Cernicharo:2012p4397}, and is so far the only detection of this species in the interstellar medium.

Here, we present observations of the formyl and methoxy radicals, as well as methanol in a sample of prestellar cores, which we supplement with H$_2$CO data from the literature in order to study the intermediate steps towards the formation of CH$_3$OH by successive hydrogenation of CO molecules on cold grain surfaces. The paper is organised as follows. In section \ref{obs}, we describe the observations, and in section \ref{results} we derive the HCO, CH$_3$O, and CH$_3$OH abundances. In section \ref{discus}, we discuss chemical routes to form HCO and CH$_3$O, both grain-surface and gas phase, and propose a likely mechanism for their formation in the cold dense gas. Finally, we conclude in section \ref{cl}.

\section{Observations}

\label{obs}
Our sample is made up of eight sources: \object{L1689B}, \object{L1495A-S}, \object{L429}, \object{L1709A}, \object{TMC2}, \object{L1521E}, \object{L1512}, and \object{L1517B}. They were chosen in order to sample different star forming regions, and physical and chemical properties. 
The source properties and coordinates are summed up in Table\,\ref{sources}. The sources were observed at the peak of their millimetre continuum emission with the IRAM 30\,m telescope, Pico Veleta, Spain, during various observing runs between January 2003 and May 2014 as described below.

\begin{table*}[htb]
\caption{Observed sources and their properties. The densities are taken from the references in Col.\,6. }
\begin{tabular}{lcclll}
\hline\hline
Source name & R.A. & Dec & n$_c({\rm H_2}$) & Region & References\\
\hline
L1689B & 16:34:48.30 & $-$24:38:04.0 & 2\;10$^5$ & Ophiuchus & (1)\\
L1495A-S & 04:18:39.90 &  $+$28:23:16.0 & 4\;10$^5$ & Taurus & (2)\\
L1709A & 16:32:45.60 &  $-$23:52:38.0 & 10$^5$ & Ophiuchus & (5)\\
L429 & 18:17:05.60  & $-$08:13:30.0 & 6\,10$^5$ & Aquila & (5)\\
TMC2 & 04:32:45.50 &  $+$24:25:08.0 & 2\;10$^4$& Taurus & (2)\\
L1521E & 04:29:14.9 & $+$26:13:57.0 & 3\;10$^5$ & Taurus & (2) \& (4) \\
L1512 & 05:04:08.6 & $+$32:43:25.0 & 8\;10$^4$ & Taurus & (2)\\
L1517B & 04:55:18.3 & $+$30:37:50.0 & 2\;10$^5$ & Taurus & (3)\\
\hline
\end{tabular}
\label{sources}
\tablefoot{References (1): \citet{Roy:2014p4845}; (2) \citet{BradyFord:2011p2416}; (3) \citet{Tafalla:2004p1027}; (4) \citet{Tafalla:2004p3983}; (5) \citet{Bacmann:2002p1238}}
\end{table*}

HCO and CH$_3$O were simultaneously observed   at 86.7\,GHz and 82.4\,GHz, respectively, with the EMIR090 receiver connected to a Fourier Transform Spectrometer (FTS) at the spectral resolution of 50\,kHz (corresponding to a velocity resolution of $\sim$\,0.18 km/s) in January 2012 for L1689B and June 2012 for the other sources.  Several methanol transitions were observed between 2011 and 2014 with the EMIR090 receiver between 80 and 110\,GHz and with the EMIR150 receiver between 143.3 and 145.15\,GHz, with the 50\,kHz FTS as backend. The complete list of transitions is given in Table\,\ref{transitions}. Since  the spectrometer has a large instantaneous bandwidth (a total of 8\,GHz can be covered non-contiguously in both polarisations in one frequency setting), many methanol transitions can be observed simultaneously. Nevertheless, only transitions with upper-level energies lower than $\sim$\,40\,K were considered, since higher energy transitions of this molecule are not excited at the low temperatures (around 10\,K) prevailing in the sources. Similarly, methanol transitions with Einstein coefficients lower than typically 10$^{-7}$\,s$^{-1}$ are intrinsically very weak. They are not detected in our sources so we do not analyse them in this study.  
Velocity resolutions ranged from $\sim 0.18$\,km/s towards 80\,GHz to about 0.10\,km/s at 145\,GHz. All data were taken in the frequency-switching mode with a frequency throw of 7.5\,MHz to reduce standing waves.
During the observations, the weather conditions were good with system temperatures close to 150\,K at 3\,mm and 250\,K at 2\,mm for the low declination sources and 80\,K at 3\,mm and 150\,K at 2\,mm for Taurus sources. Pointing was checked about every 1.5\,hrs on nearby quasars and found to be accurate within 3--4\arcsec at 3\,mm and 2-3\arcsec at 2\,mm. The beam sizes vary from 16\arcsec at 145\,GHz to 26\arcsec at 96\,GHz and 29\arcsec at 85\,GHz. The forward efficiencies were $\eta_{\rm fwd}$=0.95 at 3\,mm and $\eta_{\rm fwd}$=0.93 at 2\,mm. Beam efficiencies $\eta_{\rm beam}$ were taken to be 0.81 at 85\,GHz, 0.80 at 95\,GHz, 0.79 at 105\,GHz, and 0.74 at 145\,GHz. The data were converted from the antenna temperature $T_a^*$ to the main beam temperature scale $T_{\rm mb}$ using the values of the efficiencies given above. 

Previous CH$_3$OH observations were taken with the ABCD receivers at the IRAM 30\,m telescope in January and May 2003 for the sources L1689B, L1709A, and L429.  As backend, we used the autocorrelator VESPA with a frequency resolution of 40\,kHz at 3\,mm and 2\,mm, and 80\,kHz at 1\,mm, corresponding to velocity resolutions of 0.12\,km/s--0.08\,km/s depending on frequency. These data were taken in the position-switching mode with an off position free of methanol emission. The targeted transitions are also listed in Table\,\ref{transitions}. The main beam efficiencies were 0.76, 0.69, 0.67, 0.57, and 0.46 at 96\,GHz, 145\,GHz, 157\,GHz, 206\,GHz, and 254\,GHz, respectively, while forward efficiencies were 0.95 at 96\,GHz, 0.93 at 145 and 157\,GHz, 0.91 at 206\,GHz, and 0.88 at 254\,GHz. As with the newer observations described above, the spectra were converted to the main beam temperature scale $T_{\rm mb}$ using these values.  
During these observing runs, pointing was checked every 1.5--2 hours and was found to be accurate to within 3--4\arcsec. The beam sizes were 12\arcsec and 10\arcsec at 206\,GHz and 254\,GHz, respectively (and the same as given above at the lower frequencies). The H$_2$CO observations used in this study  for the sources L1689B, L429, and L1709A were taken from \citet{Bacmann:2003p1218}, and those for L1517B are taken from  \citet{Tafalla:2006p3988}. 

\begin{table*}
\caption{Molecular transitions observed, with their frequencies, upper-level energies E$_{\rm up}$, upper-level degeneracy g$_{\rm up}$ , Einstein spontaneous emission coefficient A$_{\rm ul}$. Quantum numbers are $N_{K_aK_c}\;J\;F$ for HCO, $J_{K}$ for CH$_3$OH, and $N_K;\Lambda;J;F$ for CH$_3$O. HCO \citep{Saito:1972p4219} and CH$_3$O \citep{Endo:1984p2512} data are taken from the JPL spectroscopic database \citep{Pickett:1998p4433}. Data for CH$_3$OH are from the CDMS catalogue \citep{Muller:2001p2418,Muller:2005p4298} and were originally determined by \citet{Xu:1997p4300} and \citet{Muller:2004p4299}}
\begin{tabular}{lcccccc}
\hline\hline
Molecule & Transition & Frequency  (MHz) & E$_{\rm up}$ (K) & g$_{\rm up}$ & A$_{\rm ul}$ (s$^{-1}$) & Notes \\ 
\hline
HCO & 1$_{0 1}$ 3/2, 2 -- 0$_{0 0}$ 1/2, 1& 86670.76 & 4.2 & 5 & 4.69 10$^{-6}$ & (1) \\
HCO & 1$_{0 1}$ 3/2, 1 -- 0$_{0 0}$ 1/2, 0 & 86708.36 & 4.2 & 3 & 4.60 10$^{-6}$ & (1) \\
HCO & 1$_{0 1}$ 3/2, 1 -- 0$_{0 0}$ 1/2, 1 & 86777.46 & 4.2 & 3 & 4.61 10$^{-6}$ & (1) \\
HCO & 1$_{0 1}$ 1/2, 0 -- 0$_{0 0}$ 1/2, 1 & 86805.78 & 4.2 & 1 & 4.71 10$^{-6}$ & (1) \\
CH$_3$O E &  1$_{ 0}$  -1  3/2, 1  --  0$_{0}$  1  3/2 , 1  &  82341.515 &  4.0 & 3 & 3.255 10$^{-6}$   & (1)\\ 
CH$_3$O E &  2$_{-1}$  -1 3/2, 1  -- 1$_{1}$  -1 1/2, 0 &   82367.891  & 17.4 & 3 & 6.60 10$^{-6}$  & (1)\\
CH$_3$O E & 2$_{1}$ -1 3/2, 1 -- 1$_{-1}$ -1 1/2, 0 &    82370.263 & 17.4 & 3 &   6.50 10$^{-6}$    & (1) \\
CH$_3$O E & 2$_{-1}$ -1 3/2, 2  -- 1$_{1}$ -1 1/2, 1 &   82398.376  & 17.4 & 5 &  9.75 10$^{-6}$  & (1) \\  
CH$_3$O E & 2$_{1}$ -1 3/2, 2 -- 1$_{-1}$ -1 1/2, 1 &   82398.879 & 17.4 & 5 &   9.75 10$^{-6}$ & (1)\\    
CH$_3$O E & 2$_{1}$ -1 3/2, 1 -- 1$_{-1}$ -1 1/2, 1 &   82409.609 & 17.4 & 3 &   3.251 10$^{-5}$ & (1)\\
CH$_3$O E & 2$_{-1}$ 1 3/2, 1 -- 1$_{1}$ -1 1/2, 1 &  82416.751 & 17.4 & 3 &  3.251 10$^{-6}$ & (1) \\
CH$_3$O E & 1$_{0}$ 1 3/2, 1 -- 0$_{0}$ -1 1/2, 0 &   82455.980 & 4.0 & 3 &  6.521 10$^{-6}$ & (1) \\
CH$_3$O E & 1$_{0}$ -1 3/2, 2 -- 0$_{0}$ -1 1/2, 1 &   82458.252 & 4.0 & 5   & 9.781 10$^{-6}$ & (1)\\    
CH$_3$O E & 1$_{0}$ -1 3/2, 2 -- 0$_{0}$ 1 1/2, 1 &  82471.825 &   4.0 & 5 & 9.783 10$^{-6}$ & (1)\\
CH$_3$O E & 1$_{0}$ -1 3/2,1 -- 0$_{0}$ 1 1/2, 0 & 82524.180 & 4.0 & 3 &  6.526 10$^{-6}$ & (1)\\    
CH$_3$O E & 1$_{0}$ 1 3/2, 1 -- 0$_{0}$ -1 1/2, 1 &   82545.726 & 4.0 & 3 &   3.264 10$^{-6}$ & (1)\\   
CH$_3$OH E & 5$_{-1}$ -- 4$_{0}$ & 84521.169 & 40.4 & 11 & 1.97 10$^{-6}$ & (2)\\
CH$_3$OH A$^+$ & 2$_{1}$ -- 1$_{1}$ & 95914.309 & 21.5 & 5 & 2.494 10$^{-6}$ & (3)\\
CH$_3$OH E & 2$_{-1}$ -- 1$_{-1}$& 96739.362 & 12.5 & 5 & 2.557 10$^{-6}$ & (3), (4)\\
CH$_3$OH A$^+$ & 2$_{0}$ -- 1$_{0}$ & 96741.375 & 7.0 & 5 & 3.408 10$^{-6}$ & (3), (4)\\
CH$_3$OH E & 2$_{0}$ -- 1$_{0}$ & 96744.550 & 20.1 & 5 & 3.408 10$^{-6}$ & (3), (4)\\
CH$_3$OH E & 2$_{1}$ -- 1$_{1}$ & 96755.511 & 28.0 & 5 & 2.624 10$^{-6}$ & (3), (4)\\
CH$_3$OH  A$^-$ & 2$_{1}$ -- 1$_{1}$ & 97582.804 & 21.6 & 5 & 2.627 10$^{-5}$ & (5) \\
CH$_3$OH A$^+$ & 3$_{1}$ -- 4$_{0}$ & 107013.803 & 28.4 & 7 & 6.131 10$^{-6}$ & (2)\\
CH$_3$OH E & 0$_{0}$ -- 1$_{-1}$ & 108893.963 & 13.1 & 1 & 1.470 10$^{-5}$ & (1)\\
CH$_3$OH A$^{+}$ & 3$_{1}$ -- 2$_{1}$ & 143865.797 & 28.4 & 7 & 1.069 10$^{-5}$ & (1)\\
CH$_3$OH E & 3$_{0}$ -- 2$_{0}$ & 145093.707 & 27.1 & 7 & 1.232 10$^{-5}$ & (4), (5), (7)\\
CH$_3$OH E & 3$_{-1}$ -- 2$_{-1}$& 145097.370 & 19.5 & 7 & 1.095 10$^{-5}$ & (4), (5), (7)\\
CH$_3$OH A$^+$ & 3$_{0}$ -- 2$_{0}$ & 145103.152 & 13.9 & 7 & 1.232 10$^{-5}$ & (4), (5), (7)\\
CH$_3$OH E & 3$_{2}$ -- 2$_{2}$ & 145126.190 & 36.2 & 7 & 6.767 10$^{-6}$ & (4), (5), (7) \\
CH$_3$OH E & 3$_{-2}$ -- 2$_{-2}$ & 145126.392 & 39.9 & 7 & 6.855 10$^{-6}$ & (4), (5), (7)\\
CH$_3$OH E & 3$_{1}$ -- 2$_{1}$& 145131.855 & 35.0 & 7 & 1.124 10$^{-5}$ & (4), (5), (7) \\
CH$_3$OH A$^{-}$ &  3$_{1}$ -- 2$_{1}$ &  146368.344 & 28.6 & 7 & 1.126 10$^{-5}$ & (7) \\
CH$_3$OH A$^{+}$ & 2$_{1}$ -- 3$_{0}$ & 156602.413 & 21.5 & 5 & 1.785 10$^{-5}$ & (6) \\
CH$_3$OH E & 4$_{0}$ -- 4$_{-1}$ & 157246.056 & 36.4 & 9 & 2.098 10$^{-5}$ & (4), (6), (7)\\
CH$_3$OH E & 1$_{0}$ -- 1$_{-1}$ & 157270.851 & 15.5 & 3 & 2.205 10$^{-5}$ & (4), (6), (7) \\
CH$_3$OH E & 3$_{0}$ -- 3$_{-1}$ & 157272.369 & 27.1 & 7 & 2.146 10$^{-5}$ & (4), (6), (7)\\
CH$_3$OH E & 2$_{0}$ -- 2$_{-1}$ & 157276.058 &  20.1 & 5 & 2.182 10$^{-5}$ & (4), (6), (7) \\
CH$_3$OH A$^+$ & 1$_{1}$ -- 2$_{0}$ & 205791.270 & 16.8 & 3 & 3.362 10$^{-5}$ & (4)\\
CH$_3$OH E & 2$_{0}$ -- 1$_{-1}$ & 254015.340 & 20.1 & 5 & 1.902 10$^{-5}$ & (4)\\
CH$_3$OH E & 2$_{1}$ -- 1$_{0}$ & 261805.710 & 28.0 & 5 & 5.572 10$^{-5}$ & (4)\\

\hline
\end{tabular}
\tablefoot{(1) Observations of all sources carried out in 2012; (2) Observations of L1689B carried out in 2012; (3) Observations of all sources  except L429 and L1709A carried out in 2011-2012; (4) Observations of L1689B, L1709A and L429 carried out in 2003; (5) Observations of all sources except L1689B carried out in 2012; (6) Observations of L1512, L1517B and L1521E carried out  in 2012; (7) Observations of L1689B carried out in 2013 and 2014.}
\label{transitions}
\end{table*}

The frequencies, upper-level energies and degeneracies, and Einstein coefficients of the observed molecular transitions are listed in Table\,\ref{transitions}). Both the methoxy radical and methanol have A- and E- type states, which were considered as separate species (though only E-CH$_3$O was observed). 
Deep integrations were performed around 82-86\,GHz, since CH$_3$O lines are weak with $rms$ noise down to 2-6\,mK for most sources. For higher frequency observations targeting the much stronger methanol lines, $rms$ noise is in the range of 30-60\,mK. 

\begin{figure}
   \centering
   \includegraphics[width=\hsize]{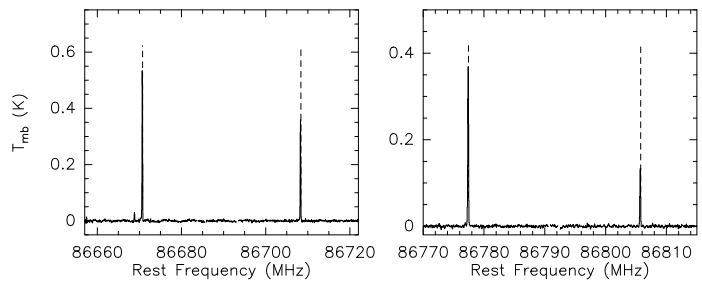}
\caption{HCO spectra in L1689B. The vertical dashed lines indicate the positions of the HCO transitions.}
\label{hcofig}
\end{figure}

\begin{figure}
   \centering
   \includegraphics[width=\hsize]{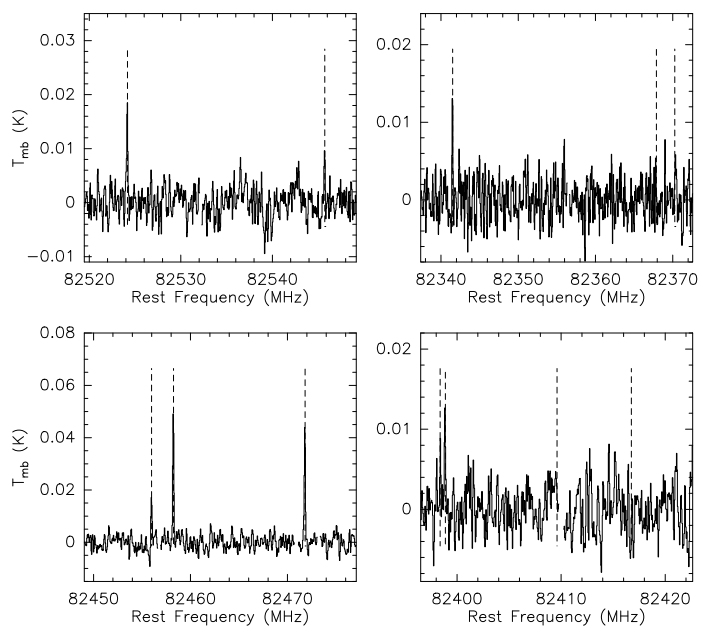}
\caption{CH$_3$O spectra in L1689B. The vertical dashed lines indicate the positions of the CH$_3$O transitions.}
\label{ch3ofig}
\end{figure}

\begin{figure*}
   \centering
          \includegraphics[width=\hsize]{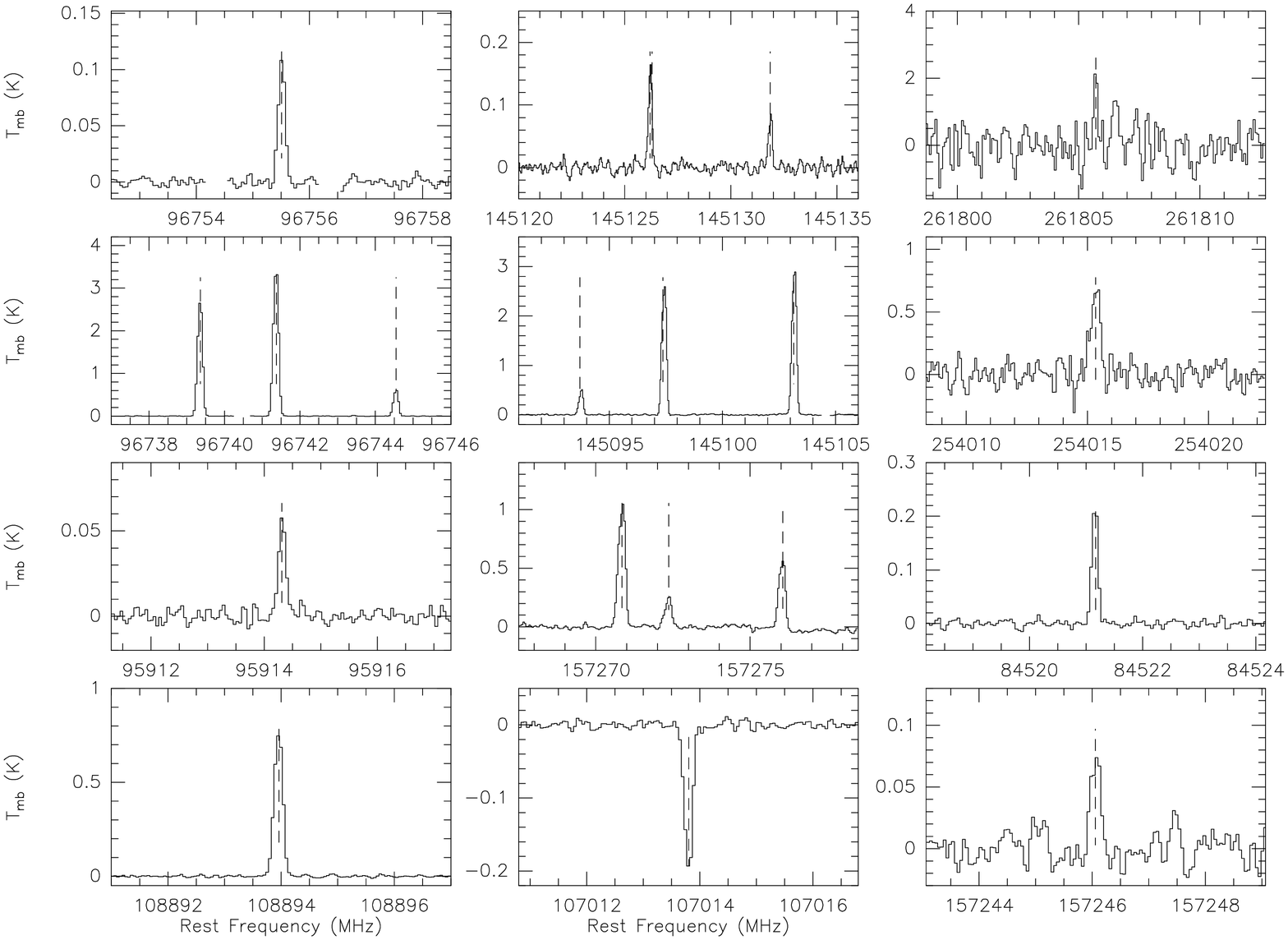}
\caption{CH$_3$OH spectra in L1689B. The vertical dashed lines indicate the positions of the CH$_3$OH transitions. Channels appearing as absorption features arising from the frequency-switching observing procedure were blanked for clarity.}
\label{ch3ohfig}
\end{figure*}

Observations were reduced using the IRAM GILDAS/CLASS software\footnote{http://www.iram.fr/IRAMFR/GILDAS}, which consisted in coadding the spectra and folding them in order to deconvolve them from the frequency-switching procedure. A low order (typically 3) polynomial was then fitted over line-free regions to correct for baseline oscillations. The resulting spectra were then converted from the $T_A^*$ to the $T_{\rm mb}$ scale using the values of the efficiencies given above. The lines were fitted with a Gaussian from which the peak temperatures, linewidths, and velocity-integrated line intensities were derived. The error on the integrated intensities was taken to be the quadratic sum of the statistic error and a calibration error, estimated to be 10\% at 3\,mm, 15\% and 2\,mm, and 20\% at 1\,mm. For non-detections, a 3\,$\sigma$ upper limit defined as 3\,$\sigma_{rms}\sqrt{\Delta v \Delta V}$ is given, where $\Delta v$ is the linewidth, $\Delta V$ the velocity resolution, and $\sigma_{rms}$ the $rms$ noise determined from the baseline fitting. The resulting line parameters are given in Tables\,\ref{tablehco}, \ref{tablech3o}, and\,\ref{tablech3oh} for HCO, CH$_3$O, and CH$_3$OH, respectively. For some sources, methanol was observed at two different epochs, and where observations are duplicate, we chose to use the newer observations in order to have a main body of observations coming from the same receivers. The older observations are only used when no other observation is available. We did check, however, that new and old observations were consistent within the uncertainties. Spectra of HCO, CH$_3$O, and CH$_3$OH in L1689B are shown in Figs.\,\ref{hcofig},\,\ref{ch3ofig}, and\,\ref{ch3ohfig}, respectively.

%

\section{Results}

\label{results}
\subsection{HCO and CH$_3$O} 

The formyl radical was detected in all sources, showing it is ubiquitous towards cold starless cores (Table\,\ref{tablehco}).  The non-detection of HCO in L183 by \citet{Schenewerk:1988p4307} can be explained by the lack of sensitivity of their observations, since in our weakest sources,  HCO lines are at 25-30\,mK, consistent with their upper limits. The methoxy radical, on the other hand, is detected in only four out of eight sources (Table\,\ref{tablech3o}), but its lines are much weaker (typically a factor of 10 less than HCO lines), and it is likely that it is present in all the sources of our sample.

There are no available collisional rates of HCO and CH$_3$O with H$_2$
or He, therefore we had to rely on an LTE-based approach to determine
the column densities and abundances of these species, assuming a
single excitation temperature $T_{\rm ex}$ for all transitions. This approximation is likely to hold since {\it i)} the hyperfine
transitions belong to the same rotational transition and {\it ii)}
opacities are low \citep[see][]{Faure:2012p5302}. In this approximation,
supposing Gaussian profiles, the  integrated line intensity $W$ is
given by \\
\begin{displaymath}
W =( J_{\nu}(T_{\rm ex}) - J_{\nu}(T_{\rm CMB})) (1-{\rm e}^{-\tau_0})\;{\rm d}\upsilon \;1/2\sqrt{\pi/\ln(2)}
\end{displaymath}
where $T_{\rm CMB}$ is the cosmic microwave background temperature, $\tau_0$ the line centre opacity, ${\rm d}\upsilon$ the linewidth, and $J_{\nu}$ refers to the radiation temperature at frequency $\nu$ defined by $J_{\nu}(T)=h\nu/k\;(\exp(h\nu/kT)-1)^{-1}$. The line centre opacity is\\
\begin{displaymath}
\tau_0 = 2 \sqrt{\ln(2)/\pi} \;N/{\rm d}\upsilon\;c^3/(8\pi\nu^3) g_{\rm up} A_{\rm ul} /Q_{\rm part} {\rm e}^{-E_u/k T_{\rm ex}} ({\rm e}^{-h\nu/k T_{\rm ex}}-1)
\end{displaymath}
with $N$ the total molecular column density, $c$ the light velocity, $g_{\rm up}$ the upper level energy, $A_{\rm ul}$ the Einstein spontaneous emission coefficient, $\nu$ the frequency of the transition, and $Q_{\rm part}$ the partition function derived at temperature $T_{\rm ex}$. 

With the above formulae, grids of  integrated intensities were calculated for a range of values in $T_{\rm ex}$ and $N$. For each value of [$T_{\rm ex}$, $N$], a $\chi^2$ was derived to assess the distance between the modelled and the observed integrated line intensities. The $\chi^2$ isocontours were plotted as a function of $T_{\rm ex}$ and $N$, 1$-\sigma$ contours corresponding to $\chi^2_{\rm min}+2.3$, as appropriate for a two-parameter model (\citet{Avni:1976p3997}.  The $\chi^2$ was also minimised for fixed values of the excitation temperature (4, 6, and 8\,K), as explained below. 

\subsubsection{HCO}

Our data do not enable us to determine the excitation temperature, because any excitation temperature above 3.5\,K would yield a fit within the 1-$\sigma$ contours; however, the plots show that for values of $T_{\rm ex}$ higher than typically 4\,K, the column density does not change significantly. For $T_{\rm ex}$ lower than $\sim$\,4\,K, the column densities cannot be determined with accuracy, because they increase quickly with small decreases in \tex,  since the opacities of the lines will increase. \citet{Gerin:2009p4072} determine an excitation temperature of 5\,K towards the PDR of the Horsehead and assume the same $T_{\rm ex}$ for their dense core position. Generally the derived $\chi^2$ values at fixed temperatures tend to favour  excitation temperatures of 4\,K over 6 or 8\,K (except for L1521E and L1517B), but most of the time the difference is small. The $\chi^2$ minimised over the whole parameter space was obtained for \tex\ values close to 4\,K (except for L1517B and L1521E for which it was 7-8\,K) and even as low as 3\,K for L429, TMC\,2 and L1512, but as before, models with other values of \tex\ were not significantly worse. Because of the impossibility of determining the excitation temperature, we chose here to determine the column densities at $T_{\rm ex}=4\,$K for all sources (Table\,\ref{coldenshco}). If the actual \tex\ is above 4\,K, this will not change the column density values very much. We cannot, however, completely exclude \tex\ values between 3 and 4\,K because such values have already been observed in other species \citep{Padovani:2011p4178,HilyBlant:2013p4440}. Should the excitation temperatures actually be below 4\,K, the column densities would be higher than those quoted in Table\,\ref{coldenshco}. Rates for collisional excitation, as well as higher upper energy level transitions, would help lift the degeneracy between \tex\ and the column density.
The column densities derived span a range of 7.1\, 10$^{11}$\,cm$^{-2}$ to 1.3\, 10$^{13}$\,cm$^{-2}$, i.e. a factor of nearly 20. This encompasses the value derived by \citet{Gerin:2009p4072} towards their dense core source and that of \citet{Cernicharo:2012p4397} towards B1-b.

\subsubsection{CH$_3$O}

For CH$_3$O, the situation is similar to HCO: determining the excitation temperature is not possible for most sources.  Above 4\,K, the column density does not depend on \tex, but below 4\,K, small variations in \tex\ lead to large uncertainties on the column densities. For L1495A-S and TMC\,2, only two lines of E-CH$_3$O were detected, and it is possible to find a model that reproduces the integrated line intensities for any assumed excitation temperature between 3\,K and 9\,K. In L1689B, on the other hand, we detected eight E-CH$_3$O lines that belong to two groups of hyperfine transitions with different upper level energies (Table\,\ref{transitions}), which enables us to constrain the excitation better. The best-fitting model is obtained for $T_{\rm ex}=8$\,K, with the best model for $T_{\rm ex}=6$\,K being clearly worse (despite the best fitting column density being similar). For this reason, we chose to determine the column densities for $T_{\rm ex}=8$\,K in all sources. This value of the excitation temperature is close to the value of  $T_{\rm ex}=10\pm 3$\,K derived by \citet{Cernicharo:2012p4397} from their data. The CH$_3$O column densities  are given in Table\,\ref{coldensch3o}. Since only E-type CH$_3$O was observed, we assumed an A/E CH$_3$O ratio of 1.25 to derive the total (A+E) CH$_3$O column density. This corresponds to the ratio of the partition functions of both subspecies at a temperature of 10\,K (close to the cores kinetic temperatures), which is expected to be the abundance ratio of the molecules if they equilibrated on grain surfaces and preserved their A/E ratio after evaporating. In the high temperature limit, the ratio of the partition functions is equal to 1. Upper limits at 3\,$\sigma$ in the sources where CH$_3$O was not detected were also calculated for $T_{\rm ex}=8$\,K. As for HCO, the column density values  will be higher if the line excitation temperature is lower than 4\,K. 

The column density values range from $\sim$ 5 10$^{11}$\,cm$^{-2}$ to 1.2 10$^{12}$\,cm$^{-2}$, although the 3$\sigma$ upper limits for L1517B, L1512, and L1521E indicate column densities lower than 2 10$^{11}$\,cm$^{-2}$. This is also consistent with the CH$_3$O column density derived by \citet{Cernicharo:2012p4397} towards B1-b.

\subsection{CH$_3$OH}

Methanol was detected in all sources. Like for the previously discussed species, HCO and CH$_3$O, the lines in L1689B are by far  the strongest of the sources in the sample. Line strengths (and detection rates) varied widely across the source sample with L1512 and L1517B showing only a few weak lines, despite the $rms$ sensitivity being similar as in most other sources. The $\rho$ Oph sources and L429 (in Aquila) have  larger linewidths ($\Delta v\sim\,$0.45\,km s$^{-1}$) than the Taurus sources, for which $\Delta v\sim$\,0.35\,km s$^{-1}$. This is generally attributed to a more turbulent environment in $\rho$ Oph. Between 6 and 19 transitions were detected both in A- and in E- CH$_3$OH depending on the source (Table\,\ref{tablech3oh}). 
 
Line parameters (linewidth, integrated line intensity, peak line intensity) are given in Table\,\ref{tablech3oh}. 
Interestingly, the A-CH$_3$OH line at 107.013\,GHz is seen in absorption against the cosmic microwave background. This is similar to the absorption seen at cm wavelengths in H$_2$CO \citep{Palmer:1969p4000,Townes:1969p5328} or in SO$_2$   \citep{Cernicharo:2012p5353}. In CH$_3$OH, the absorption occurs at relatively high densities (a few 10$^{5}$\,cm$^{-3}$) and brings strong constraints on the physical conditions prevailing in prestellar cores. We postpone the discussion of this aspect to a future paper.

The column densities for CH$_3$OH were determined using the RADEX radiative transfer code \citep{vanderTak:2007p4401} using the large velocity gradient (LVG) approximation. The rate coefficients for collisional excitation of CH$_3$OH with H$_2$ used in our LVG calculation were computed by \citet{Rabli:2010p4127}. Only collisions with para-H$_2$ were considered in our calculation because ortho-H$_2$ represents
at most 0.1-1\% of the total H$_2$ in prestellar cores, where the gas temperature is around 10\,K  \citep{Faure:2013p4435,Pagani:2009p1030,Dislaire:2012p4434}. We did, however, check that adding 1\% of ortho-H$_2$ does not change the outcome of our calculation. No a priori A-CH$_3$OH/E-CH$_3$OH was assumed, and A- and E-type CH$_3$OH were treated as two distinct species. As methanol is likely formed
for the most part on grain surfaces \citep{Geppert:2006p4196}, the A-CH$_3$OH/E-CH$_3$OH ratio is expected to be equal to the ratio of the partition functions, i.e. 1.4 at 10\,K, and decreasing towards 1 with increasing temperature. 

For each source, we calculated line intensities for ranges of values in column density, temperature, and H$_2$ density. The LVG code requires the linewidth as an input parameter, which was taken for each source as an average value of the detected lines (typically around 0.35\,km/s for the Taurus sources and 0.45\,km/s for the Rho Oph sources, see above). The linewidth, however, only has very little effect on the output integrated line intensities. The column densities were varied typically from 10$^{13}$\,cm$^{-2}$ to 2.5\,10$^{14}$\,cm$^{-2}$ in steps of 10$^{13}$\,cm$^{-2}$ (or from 10$^{12}$\,cm$^{-2}$ to 2\,10$^{13}$\,cm$^{-2}$ in steps of 10$^{12}$\,cm$^{-2}$ for L1512 and L1517B which have lower CH$_3$OH column densities), the temperatures from 6\,K to 14\,K in steps of 0.5\,K and the densities from 10$^{4}$\,cm$^{-3}$ to 10$^{6}$\,cm$^{-3}$ in steps of 10$^{4}$\,cm$^{-3}$. Finer grids in column density were used around the value best reproducing the observations, or the grids were extended when the solution was close to or at the boundary of the parameter space. The A- and E-type methanols were assumed to be coexistent; i.e., the models for A-CH$_3$OH and E-CH$_3$OH were calculated assuming the same density and temperature (but different column densities in A-CH$_3$OH and E-CH$_3$OH). Under this hypothesis, the $\chi^2$ values of the model are calculated by summing over all detected methanol lines, i.e.
\begin{eqnarray*}
\chi^2(N_A,N_E,n,T_k) & = & \sum_i\frac{(W^{\rm obs}_i - W^{\rm calc}_i(N_e,n,T_k))^2}{(\sigma^{\rm obs}_i)^2}\\
                     &   &  +\sum_j\frac{(W^{\rm obs}_j - W^{\rm calc}_j(N_a,n,T_k))^2}{(\sigma^{\rm obs}_j)^2}
\end{eqnarray*}
where the index $i$ refers to the E-CH$_3$OH lines and the index $j$ to the A-CH$_3$OH lines, $W^{\rm obs}$ and $\sigma^{\rm obs}$ are, respectively, the observed integrated intensity of the considered line and the 1-$\sigma$ error on this value (Table\,\ref{tablech3oh}), and $ W^{\rm calc}$ is the integrated intensity  calculated by the LVG model.

The E-CH$_3$OH transition at 108.895\,GHz was consistently poorly fitted by the models. In fact, the collisional rate for this line turns out to be set to zero, as is the case for a few transitions in the coupled states approximation used by \citet{Rabli:2010p4127} in their calculations. We therefore chose to exclude the line of E-CH$_3$OH at 108.894\,GHz in deriving the $\chi^2$ of the model because its excitation might not be calculated well by the LVG code. 

Some transitions (e.g. the lines at 157.271\,GHz and 254.015\,GHz) are also not very well reproduced by the model
in most of the sources. The simplicity of our model (single-density and single-temperature) probably accounts for the discrepancy between the observed and calculated integrated intensities for these lines.  
However, that our model systematically under-reproduces the intensities of these lines, which are mostly excited at high densities, is an indication that there is some methanol emission arising from higher density regions than that of our simple model.

The parameters minimising the $\chi^2$ and the derived CH$_3$OH column densities are listed in Table\,\ref{coldensch3oh}. This should yield the model that reproduces the observations
best. The 1-$\sigma$ confidence intervals on these parameters were determined from the range of model parameters verifying  $\chi^2 =  \chi^2_{\rm min}+3.5$ \citep{Avni:1976p3997}, as suited for a three-parameter model. The 1-$\sigma$ uncertainties are given in Table\,\ref{1sigmach3oh}. 

The total CH$_3$OH column densities span about a factor of 10 from 10$^{13}$\,cm$^{-2}$ in the weakest Taurus sources to 1.3\,10$^{14}$\,cm$^{-2}$ in L1689B. Other measurements of methanol abundances in dark cores \citep{Friberg:1988p3131,Guzman:2013p4399,Bizzocchi:2014p5365}  yielded column densities close to 2\,10$^{13}$\,cm$^{-2}$, which is within the range we derived, although most of our values are higher. One source of the sample, L1517B, was already observed in CH$_3$OH by \citet{Tafalla:2006p3988}, whose model is consistent with a column density of 4\,10$^{12}$\,cm$^{-2}$, when our value is nearly three times more. The presence of such a large discrepancy between our column density and that of \citet{Tafalla:2006p3988} is not completely clear. Our spectra of the lines at 96.739, 96.741, 145.097, and 145.103\,GHz, which  were taken at the continuum peak of the source, present different intensities to those of \citet{Tafalla:2006p3988}, which were averaged from several spectra in a 20$^{\prime\prime}$ radius around that peak. Whereas we estimate the column density at one position, \citet{Tafalla:2006p3988} assume an H$_2$ density profile  to fit an abundance to their methanol maps data, and it is unclear how this assumption will affect their total column density. 
Nevertheless, an important aspect in our study is that we proceed similarly and consistently  for all of our sources  and that all the species were observed at the same position for each source, to allow us to compare the various relative abundances. We therefore keep our value for the methanol column density in L1517B.   

The A/E ratios are between 1 and 1.5, with most of them around 1.2, except for L1495A-S, where it is lower than 1 (0.7). It should be noted that this is consistent with grain surface formation and nuclear-spin equilibration at temperatures around 10\,K. For L1495A-S, a ratio of 1 is still within the uncertainties. The densities optimizing the fit are above 10$^{5}$\,cm$^{-3}$ (see Table~10), which is similar to values usually derived at the centres of prestellar cores, and this would tend to indicate that some methanol emission arises from the central parts of the cores. This is a little surprising, because methanol, like most other C-bearing species, is expected to be depleted at the core centres, so that the emission should come from less dense regions \citep{Vastel:2014p5375}. Most of our density values have  large uncertainties, but  except for  TMC\,2 and L1517B, those uncertainties point towards higher values of the densities. Density estimates from the literature come from modelling of dust continuum emission and relies on assumptions on the dust emission coefficients that are still poorly constrained, so that these values are also uncertain by a factor of at least 2--3. Given these facts, the densities found here are generally in agreement with the central densities quoted in the literature for those cores, i.e. 2 10$^5$ in L1517B \citep{Tafalla:2006p3988}, 2.7 10$^{5}$cm$^{-3}$ in L1521E \citep{Tafalla:2004p3983}, 3 10$^5$ in TMC\,2, 1.5 10$^5$ in L1689B, and 6 10$^5$ in L429 \citep{Crapsi:2005p30}. The gas kinetic temperature values from our model are below 10\,K (Table\,\ref{coldensch3oh}), which is consistent with the emission coming from the denser parts of the cores. Some values are very low (6\,K), but because of the high uncertainties, this result might not be meaningful. \citet{Padovani:2011p4178}, however, do not exclude such very low temperatures in TMC\,2 and L1521E. Extreme caution should be taken interpreting densities from LVG modelling because they will be dependent on the transitions used and their critical densities. Indeed, the various transitions probed can arise from regions of different densities (and temperatures), which is not taken into account by uniform LVG models, so that the densities yielding the best fit in an LVG model should rather represent an average density in the source. A non-local radiative transfer calculation should be performed to determine the spatial origin of the emission, which, however, requires having a map of the source in several transitions and a source model. Such an analysis is beyond the scope of the current paper.

Our modelling also enables us to determine the opacities of the CH$_3$OH lines. The transitions at 96.739, 96.741, 145.103, 157.271, and 157.276\,GHz are often (moderately) optically thick, with opacities between one and two for the most optically thick ones, so that care should be taken when deriving column densities using only (some of) these lines.

\begin{table*}
\caption{Line parameters for HCO. The line intensity $T_{\rm mb}$, the linewidth $\Delta v$, and the integrated intensity were determined from Gaussian fits to the line. The error is the quadratic sum of the statistical fit error and a calibration error taken as 10\% at 3\,mm.}
\label{tablehco}
\begin{tabular}{lcccccc}
\hline\hline
Source & Frequency  & rms &  $T_{\rm mb}$  & $\Delta v$ & Integrated intensity& Error on int. intensity \\
               & MHz           & mK &     mK                     &   km s$^{-1}$ & K km s$^{-1}$ & K km s$^{-1}$\\
\hline
L1689B    & 86670.76 & 2.6 & 545 & 0.50 & 0.290 & 0.029\\
                & 86708.36 & 2.4 & 360 & 0.51 & 0.196 & 0.020\\
                & 86777.46 & 2.2 &  360 & 0.50 & 0.192 & 0.019\\
                & 86805.78 & 2.1 &  145 & 0.48 & 0.073 & 0.008\\
\hline
L1709A    & 86670.76 & 3.7  & 265 & 0.50 & 0.141 & 0.014\\
                & 86708.36 & 2.8 & 170 & 0.55 & 0.100 & 0.010\\
                & 86777.46 & 3.1 &  165 & 0.51 & 0.089 & 0.009\\
                & 86805.78 & 3.3 & 55 & 0.52 & 0.031 & 0.004\\
\hline
L429        & 86670.76 & 4.6 &  140 & 0.50 & 0.075 & 0.008\\
                & 86708.36 & 4.1 &  100 & 0.65 & 0.069 & 0.007\\
                & 86777.46 & 4.8  & 95 & 0.51 & 0.050 & 0.006\\
                & 86805.78 & 4.9 & 35 & 0.46 & 0.018 & 0.003\\
\hline
L1495A-S    & 86670.76 & 4.5 & 350 & 0.37 & 0.155 & 0.016\\
                & 86708.36 & 5.5 & 230 & 0.47 & 0.123 & 0.013\\
                & 86777.46 & 4.7 & 235 & 0.40 & 0.110 & 0.011\\
                & 86805.78 & 4.6 & 95 & 0.37 & 0.041 & 0.004\\
\hline
TMC2       & 86670.76 & 5.6 & 155 & 0.38 & 0.063 & 0.007\\
                & 86708.36 & 5.6 & 100 & 0.55 & 0.057 & 0.006\\
                & 86777.46 & 5.1 &  90 & 0.37 & 0.035 & 0.004\\
                & 86805.78 & 5.3 & 35 & 0.35 & 0.013 & 0.002\\
\hline
L1521E  & 86670.76 & 3.7 &  120 & 0.46 & 0.059 &  0.006\\
                & 86708.36 & 3.4 &  285  & 0.50 & 0.151 &  0.015 \\
                & 86777.46 & 3.1 & 75   & 0.50 & 0.039 & 0.004\\
                & 86805.78 & 2.9 &  20  & 0.41 & 0.009 &  0.002\\
\hline
L1512   & 86670.76 & 4.7 & 70  & 0.36 & 0.026 &  0.003 \\
                & 86708.36 & 4.4 &  40 & 0.48 & 0.020 &   0.003 \\
                & 86777.46 & 4.1 &  50 & 0.31 & 0.016 &  0.002\\
                & 86805.78 & 4.6 & $-$ & $-$ & $<0.003$  & $-$ \\
\hline
L1517B  & 86670.76 & 4.2 &  55 & 0.41 & 0.024 &  0.003 \\
                & 86708.36 & 4.6 &  25 & 0.38 & 0.011 &  0.002\\
                & 86777.46 & 4.0 & 30   & 0.35 & 0.012 & 0.002\\
                & 86805.78 & 4.8 & $-$ & $-$ & $<0.004$  & $-$ \\
\hline
\end{tabular}
\tablefoot{The HCO line at 86708.36\,MHz in L1521E is blended with the $J:15-14$ C$_3$S line at 86708.38\,MHz, which explains that for this source, the relative intensities of the HCO lines deviate strongly from the statistical 5:3:3:1 ratios expected in the optically thin case. The 86708.36\,MHz line was therefore excluded from the fit in L1521E.}
\end{table*}

\begin{longtab}
\begin{longtable}{lcccccc}
\caption{\label{tablech3o}Line parameters for CH$_3$O. The line intensity $T_{\rm mb}$, the linewidth $\Delta v$, the integrated intensity were determined from Gaussian fits to the line. The error is the quadratic sum of the statistical fit error and a calibration error taken as 10\% at 3\,mm.}\\
\hline\hline
Source & Frequency & rms &  $T_{\rm mb}$ & $\Delta v$  & Integrated intensity & Error on int. intensity \\
               & MHz           & mK &     mK                     &   km s$^{-1}$ & K km s$^{-1}$ & K km s$^{-1}$\\
\hline
\endfirsthead
\caption{continued.}\\
\hline
Source & Frequency  & rms &  $T_{\rm mb}$  & $\Delta v$ & Integrated intensity& Error on int. intensity \\
               & MHz           & mK &     mK                     &   km s$^{-1}$ & K km s$^{-1}$ & K km s$^{-1}$\\
\hline
\endhead
\hline
\endfoot
L1689B & 82341.52 & 2.7 & 16 & 0.42 & 7.2 10$^{-3}$ & 1.3 10$^{-3}$ \\
              & 82367.89 & 2.1 &  $-$ & $-$ & $<$ 1.9 10$^{-3}$ & $-$ \\
              & 82370.26 & 2.4 &  $-$ & $-$ & $< $ 2.1 10$^{-3}$ & $-$ \\
              & 82398.38 & 2.7 & 10 & 0.36 & 3.7 10$^{-3}$ & 1.1 10$^{-3}$ \\
              & 82398.88 & 2.6 & 13 & 0.49 & 6.7 10$^{-3}$ & 1.3 10$^{-3}$ \\
              & 82409.61 & 2.8 &  $-$ & $-$ & $< $ 2.5 10$^{-3}$ & $-$\\
              & 82416.75 & 2.5 &  $-$ & $-$ & $< $ 2.3 10$^{-3}$ & $-$\\
              & 82455.98 & 2.5 & 19 & 0.48 & 9.6 10$^{-3}$ & 1.4 10$^{-3}$\\
              & 82458.25 & 2.7 & 50 & 0.48 & 25.4 10$^{-3}$ & 2.8 10$^{-3}$\\
              & 82471.83 & 2.5 & 45 & 0.54 & 26.5 10$^{-3}$ & 2.9 10$^{-3}$\\
              & 82524.18 & 2.6 & 18 & 0.58 & 11.1 10$^{-3}$ & 1.7 10$^{-3}$ \\
              & 82545.73 & 2.3 & 11 & 0.50 & 6.0 10$^{-3}$ & 1.2 10$^{-3}$\\
\hline
L1709A & 82341.52 & 4.3 & $-$ & $-$ & $<$  3.4 10$^{-3}$ & $-$\\
              & 82367.89 & 3.3 & $-$ & $-$ & $<$  2.7 10$^{-3}$ & $-$ \\
              & 82370.26 & 3.9 & $-$ & $-$ & $<$  3.2 10$^{-3}$ & $-$\\
              & 82398.38 & 3.8 & $-$ & $-$ & $<$  3.1 10$^{-3}$ & $-$\\
              & 82398.88 & 3.9 & $-$ & $-$ & $<$  3.1 10$^{-3}$ & $-$\\
              & 82409.61 & 4.1& $-$ & $-$ & $<$  3.3 10$^{-3}$ & $-$\\
              & 82416.75 & 3.8 & $-$ & $-$ & $<$  3.0 10$^{-3}$ & $-$\\
              & 82455.98 & 4.3 & $-$ & $-$ & $< $ 3.4 10$^{-3}$ & $-$\\
              & 82458.25 & 4.2 & 19 & 0.43 & 8.6 10$^{-3}$ & 1.9 10$^{-3}$\\
              & 82471.83 & 4.0 & 25 & 0.48 &13.9 10$^{-3}$ & 2.3 10$^{-3}$\\
              & 82524.18 & 3.6 & 10 & 0.48 & 5.2 10$^{-3}$ & 1.7 10$^{-3}$\\
              & 82545.73 & 4.1 & 13 & 0.55 & 6.7 10$^{-3}$ & 2.1 10$^{-3}$\\
\hline
L429      & 82341.52 & 9.5 & $-$ & $-$  & $<$  8.2 10$^{-3}$ & $-$\\
              & 82367.89 & 11.0 & $-$ & $-$ & $<$  9.6 10$^{-3}$ & $-$\\
              & 82370.26 & 10.5 & $-$ & $-$ & $<$  9.3 10$^{-3}$ & $-$\\
              & 82398.38 & 9.0 & $-$ & $-$ & $<$  7.9 10$^{-3}$ & $-$\\
              & 82398.88 & 9.4 & $-$ & $-$ & $<$  8.2 10$^{-3}$ & $-$\\
              & 82409.61 & 10.5 & $-$ & $-$ & $<$  9.2 10$^{-3}$ & $-$\\
              & 82416.75 & 9.8 & $-$ & $-$ & $<$  8.6 10$^{-3}$ & $-$\\
              & 82455.98 & 10.3 & $-$ & $-$ & $<$  9.2 10$^{-3}$ & $-$\\
              & 82458.25 & 10.8 & $-$ & $-$ & $<$  9.4 10$^{-3}$ & $-$\\
              & 82471.83 & 9.1 & $-$ & $-$ & $<$  8.0 10$^{-3}$ & $-$\\
              & 82524.18 & 10.3 & $-$ & $-$ & $<$  9.1 10$^{-3}$ & $-$\\
              & 82545.73 & 9.2 & $-$ & $-$ & $<$  8.1 10$^{-3}$ & $-$\\
\hline
L1495A-S & 82341.52 & 5.5 & $-$ & $-$ & $<$  4.4 10$^{-3}$ & $-$\\
              & 82367.89 & 6.0 & $-$ & $-$ & $<$  4.8 10$^{-3}$ & $-$\\
              & 82370.26 & 6.0 & $-$ & $-$ & $<$  4.8 10$^{-3}$ & $-$\\
              & 82398.38 & 7.5 & $-$ & $-$ & $<$  6.0 10$^{-3}$ & $-$\\
              & 82398.88 & 6.2 & $-$ & $-$ & $<$  4.9 10$^{-3}$ & $-$\\
              & 82409.61 & 5.9 & $-$ & $-$ & $<$  4.7 10$^{-3}$ & $-$\\
              & 82416.75 & 5.3 & $-$ & $-$ & $<$  4.2 10$^{-3}$ & $-$\\
              & 82455.98 & 5.5 & $-$ & $-$ & $<$  4.4 10$^{-3}$ & $-$\\
              & 82458.25 & 5.4 & 50 & 0.38 & 18.9 10$^{-3}$ & 2.8 10$^{-3}$\\
              & 82471.83 & 4.9 & 35 & 0.49 & 19.4 10$^{-3}$ & 3.0 10$^{-3}$\\
              & 82524.18 & 6.5 & $-$ & $-$ & $<$  5.2 10$^{-3}$ & $-$\\
              & 82545.73 & 4.3 & $-$ & $-$ & $<$  3.5 10$^{-3}$ & $-$\\
\hline
TMC2    & 82341.52 & 5.8 & $-$ & $-$ & $<$  4.7 10$^{-3}$ & $-$\\
              & 82367.89 & 6.2 & $-$ & $-$ & $<$  4.9 10$^{-3}$ & $-$\\
              & 82370.26 & 6.7 & $-$ & $-$ & $<$  5.4 10$^{-3}$ & $-$\\
              & 82398.38 & 7.0 &  $-$ & $-$ & $<$  5.6 10$^{-3}$ & $-$\\
              & 82398.88 & 6.9 & $-$ & $-$ & $<$  5.5 10$^{-3}$ & $-$\\
              & 82409.61 & 6.7 & $-$ & $-$ & $<$  5.4 10$^{-3}$ & $-$\\
              & 82416.75 & 7.1 & $-$ & $-$ & $<$  5.7 10$^{-3}$ & $-$\\
              & 82455.98 & 6.6 &  $-$ & $-$ & $<$  5.3 10$^{-3}$ & $-$\\
              & 82458.25 & 6.5 & 18 & 0.59 & 11.3 10$^{-3}$ & 3.7 10$^{-3}$\\
              & 82471.83 & 5.8 & 19 & 0.67 & 13.5 10$^{-3}$ & 3.7 10$^{-3}$\\
              & 82524.18 & 6.3 & $-$ & $-$ & $<$  5.1 10$^{-3}$ & $-$\\
              & 82545.73 & 6.0 & $-$ &  $-$ & $<$  4.8 10$^{-3}$ & $-$\\
\hline
L1521E  & 82341.52 & 4.1 & $-$ & $-$  & $<$  3.2 10$^{-3}$ & $-$\\
              & 82367.89 & 5.5 & $-$ & $-$ & $<$  4.3 10$^{-3}$ & $-$\\
              & 82370.26 & 5.2 & $-$ & $-$ & $<$  4.1 10$^{-3}$ & $-$\\
              & 82398.38 & 4.6 & $-$ & $-$ & $<$  3.6 10$^{-3}$ & $-$\\
              & 82398.88 & 4.6 & $-$ & $-$ & $<$  3.6 10$^{-3}$ & $-$\\
              & 82409.61 & 5.2 & $-$ & $-$ & $<$  4.0 10$^{-3}$ & $-$\\
              & 82416.75 & 4.7 & $-$ & $-$ & $<$  3.6 10$^{-3}$ & $-$\\
              & 82455.98 & 5.0 & $-$ & $-$ & $<$  3.9 10$^{-3}$ & $-$\\
              & 82458.25 & 5.7 & $-$ & $-$ & $<$  4.4 10$^{-3}$ & $-$\\
              & 82471.83 & 5.8 & $-$ & $-$ & $<$  4.6 10$^{-3}$ & $-$\\
              & 82524.18 & 5.5 & $-$ & $-$ & $<$  4.3 10$^{-3}$ & $-$\\
              & 82545.73 & 4.8 & $-$ & $-$ & $<$  3.7 10$^{-3}$ & $-$\\
\hline
L1512  & 82341.52 & 6.3 & $-$ & $-$  & $<$  4.9 10$^{-3}$ & $-$\\
              & 82367.89 & 6.2 & $-$ & $-$ & $<$  4.9 10$^{-3}$ & $-$\\
              & 82370.26 & 5.6 & $-$ & $-$ & $<$  4.3 10$^{-3}$ & $-$\\
              & 82398.38 & 6.7 & $-$ & $-$ & $<$  5.2 10$^{-3}$ & $-$\\
              & 82398.88 & 6.7 & $-$ & $-$ & $<$  5.2 10$^{-3}$ & $-$\\
              & 82409.61 & 4.8 & $-$ & $-$ & $<$  3.7 10$^{-3}$ & $-$\\
              & 82416.75 & 6.4 & $-$ & $-$ & $<$  5.0 10$^{-3}$ & $-$\\
              & 82455.98 & 6.4 & $-$ & $-$ & $<$  5.0 10$^{-3}$ & $-$\\
              & 82458.25 & 6.0 & $-$ & $-$ & $<$  4.7 10$^{-3}$ & $-$\\
              & 82471.83 & 5.7 & $-$ & $-$ & $<$  4.4 10$^{-3}$ & $-$\\
              & 82524.18 & 6.8 & $-$ & $-$ & $<$  5.3 10$^{-3}$ & $-$\\
              & 82545.73 & 6.2 & $-$ & $-$ & $<$  4.8 10$^{-3}$ & $-$\\
\hline
L1517B  & 82341.52 & 5.7 & $-$ & $-$  & $<$  4.4 10$^{-3}$ & $-$\\
              & 82367.89 & 6.5 & $-$ & $-$ & $<$  5.1 10$^{-3}$ & $-$\\
              & 82370.26 & 5.5 & $-$ & $-$ & $<$  4.3 10$^{-3}$ & $-$\\
              & 82398.38 & 6.8 & $-$ & $-$ & $<$  5.3 10$^{-3}$ & $-$\\
              & 82398.88 & 6.8 & $-$ & $-$ & $<$  5.3 10$^{-3}$ & $-$\\
              & 82409.61 & 5.7 & $-$ & $-$ & $<$  4.4 10$^{-3}$ & $-$\\
              & 82416.75 & 7.0 & $-$ & $-$ & $<$  5.4 10$^{-3}$ & $-$\\
              & 82455.98 & 4.6 & $-$ & $-$ & $<$  3.6 10$^{-3}$ & $-$\\
              & 82458.25 & 5.6 & $-$ & $-$ & $<$  4.4 10$^{-3}$ & $-$\\
              & 82471.83 & 5.9 & $-$ & $-$ & $<$  4.6 10$^{-3}$ & $-$\\
              & 82524.18 & 5.5 & $-$ & $-$ & $<$  4.3 10$^{-3}$ & $-$\\
              & 82545.73 & 5.6 & $-$ & $-$ & $<$  4.3 10$^{-3}$ & $-$\\
\hline
\end{longtable}
\end{longtab}

\begin{longtab}
\begin{longtable}{lcccccc}
\caption{\label{tablech3oh}Line parameters for CH$_3$OH. The line intensity $T_{\rm mb}$, the linewidth $\Delta v$, the integrated intensity were determined from Gaussian fits to the line. The error is the quadratic sum of the statistical fit error and a calibration error taken as 10\% at 3\,mm., 15\% at 2\,mm and 20\% at 1\,mm.}\\
\hline\hline
Source & Frequency  & rms &  $T_{\rm mb}$  & $\Delta v$ & Integrated intensity& Error on int. intensity \\
               & MHz           & mK &     K                     &   km s$^{-1}$ & K km s$^{-1}$ & K km s$^{-1}$\\
\hline
\endfirsthead
\caption{continued.}\\
\hline
Source & Frequency  & rms &  $T_{\rm mb}$  & $\Delta v$ & Integrated intensity& Error on int. intensity \\
               & MHz           & mK &     K                     &   km s$^{-1}$ & K km s$^{-1}$ & K km s$^{-1}$\\
\hline
\endhead
\hline
\endfoot
L1689B & 84521.17 & 6.2 & 230 & 0.45 & 0.109 & 0.011\\
                & 95914.31 & 3.5 & 60 & 0.48 & 0.030 & 0.003 \\
                & 96739.36 & 2.9 & 2745 & 0.48 & 1.392 & 0.14\\
                & 96741.38 & 3.0 & 3515 & 0.49 & 1.834 & 0.18\\
                & 96744.55 & 2.8 & 620 & 0.45 & 0.296 & 0.030\\
                & 96755.51 & 2.9 & 110 & 0.44 & 0.051& 0.005\\
                & 107013.80 & 4.9 & -200 & 0.46 & -0.099 & 0.010\\
                & 108893.96 & 3.1 & 785 & 0.49 & 0.409 & 0.082\\
                & 143865.80 & 45 & $-$ & $-$ & $<0.028$& $-$\\
                & 145093.71 & 7.4 & 515 & 0.43 & 0.232 & 0.035\\
                & 145097.37 & 7.8 & 2680 & 0.47 & 1.352 & 0.24\\
                & 145103.15 & 7.3 & 3025 & 0.48 & 1.549 & 0.232 \\
                & 145126.19 & 7.1 & 165 & 0.41 & 0.071 & 0.011 \\
                & 145126.39 & 7.1 & $-$ & $-$ & $< 0.004 $& $-$\\
                & 145131.85 & 6.8 & 85 & 0.41 & 0.036 & 0.006\\
                & 146368.34 & 8.8 & 65 & 0.40 & 0.026 & 0.005\\
                & 157246.06 & 11 & 80 & 0.39 & 0.033 & 0.006\\
                & 157270.85 & 12 & 1060 & 0.50 & 0.560 & 0.084\\
                & 157272.38 & 13 & 250 & 0.46 & 0.123 & 0.019\\
                & 157276.06 & 13 & 615 & 0.48 & 0.315 & 0.047\\
                & 205791.27 & 222 & $-$ & $-$ & $<0.146$ & $-$\\
                & 254015.34 & 82 & 695 & 0.52 & 0.382 & 0.081\\
                & 261805.71 & 505 & $-$ & $-$ & $<0.272$ & $-$ \\
\hline
L1709A  & 96739.36 & 43 & 1925 & 0.47 & 0.963 & 0.098\\
                & 96741.38 & 44 & 2570 & 0.46 & 1.259 & 0.127\\
                & 96744.55 & 58 & 310 & 0.37 & 0.121 & 0.022\\
                & 96755.51 & 48 & $-$ & $-$ & $<0.032$ & $-$\\
                & 97582.80 & 5.5 & 32 & 0.62 & 0.021 & 0.003\\
                & 108893.96 & 9.8 & 450 & 0.47 & 0.226 & 0.023\\
                & 143865.80 & 31 & 110 & 0.29 & 0.033 & 0.009\\
                & 145093.71 & 30 & 260 & 0.32 & 0.088 & 0.015\\
                & 145097.37 & 35 & 1880 & 0.43 & 0.860 & 0.130\\
                & 145103.15 & 36 & 2055 & 0.44 & 0.963 & 0.097\\
                & 145126.19 & 36 & 120 & 0.34 & 0.043 & 0.012\\
                & 145126.39 & 36 & $-$ & $-$ & $<0.022$ &$-$\\
                & 145131.85 & 34 & $-$ & $-$ & $<0.021$ & $-$\\
                & 157246.06 & 82 & $-$ & $-$ & $<0.043$ & $-$\\
                & 157270.85 & 87 & 735 & 0.47 & 0.368 & 0.061\\
                & 157272.38 & 83 & $-$ & $-$ & $<0.044$ & $-$\\
                & 157276.06 & 88 & 360 & 0.38 & 0.145 & 0.030\\
                & 205791.27 & 133 & $-$ & $-$ & $<0.087$ & $-$\\
                & 254015.34 & 106 & $-$ & $-$ & $<0.062$ & $-$\\
                & 261805.71 & 334 & $-$ & $-$ & $<0.194$ & $-$\\
\hline
L429            & 96739.36 & 48 & 1470 & 0.42 & 0.657 & 0.068\\
                & 96741.38 & 51 & 1995 & 0.42 & 0.892 & 0.091 \\
                & 96744.55 & 48 & 185 & 0.60 & 0.117 & 0.024\\
                & 96755.51 & 45 & $-$ & $-$ & $<0.030$ & $-$\\
                & 97582.80 &  6.6  & $-$ & $-$ & $<0.005$ &$-$\\
                & 108893.96 & 10 & 315 & 0.49 & 0.165 & 0.017\\
                & 143865.80 & 23 & $-$ & $-$ & $<0.014$ & $-$\\
                & 145093.71 & 24 & 125 & 0.43 & 0.058 & 0.012\\
                & 145097.37 & 24 & 1300 & 0.41 & 0.567 & 0.085\\
                & 145103.15 & 32 & 1410 & 0.42 & 0.630 & 0.095\\
                & 145126.19 & 25 & 90 & 0.31 & 0.030 & 0.008 \\
                & 145126.39 & 25 & $-$ & $-$ & $<0.016$ & $-$\\
                & 145131.85 & 24 & $-$ & $-$ & $<0.015$ & $-$\\
                & 157246.06 & 39 & $-$ & $-$ & $<0.021$ & $-$\\
                & 157270.85 & 43 & 460 & 0.51 & 0.250 & 0.039\\
                & 157272.38 & 44 & $-$ & $-$ & $<0.023$ & $-$\\
                & 157276.06 & 42 & 190 & 0.59 & 0.121 & 0.022 \\
                & 205791.27 & 112 & $-$ & $-$ & $<0.073$ & $-$\\
                & 254015.34 & 47& 220 & 0.51 & 0.119 & 0.028\\
                & 261805.71 & 245 & $-$ & $-$ & $<0.142$ & $-$\\
\hline
L1495AS  & 95914.31 & 7.5 & 60 & 0.26 & 0.016 & 0.003\\
                & 96739.36 & 13.2 & 2030 & 0.37 & 0.800 & 0.080\\
                & 96741.38 & 12.3 & 2425 & 0.39 & 1.006 & 0.100\\
                & 96744.55 & 10.1 & 665 & 0.33 & 0.234 & 0.024\\
                & 96755.51 & 9.1 & 110 & 0.38 & 0.045 & 0.006\\
                & 97582.80 & 3.0 & 70 & 0.37 & 0.027 & 0.003\\
                & 108893.96 & 5.9 & 530 & 0.36 & 0.205 & 0.021\\
                & 143865.80 & 21 & 55 & 0.32 & 0.019 & 0.007\\
                & 145093.71 & 23 & 755 & 0.26 & 0.209 & 0.032\\
                & 145097.37 & 25 & 2495 & 0.33 & 0.876 & 0.132\\
                & 145103.15 & 24 & 2740 & 0.33 & 0.962 & 0.145\\
                & 145126.19 & 24 & 185 & 0.27 & 0.053 & 0.011\\
                & 145126.39 & 24 & $-$ & $-$ & $<0.013$ & $-$\\
                & 145131.85 & 25 & 115 &0.36 & 0.045 & 0.010\\
\hline
TMC2      & 95914.31 & 8.5 & $-$ & $-$ & $<0.005$ & $-$\\
                & 96739.36 & 6.5 & 1015 & 0.39 & 0.422 & 0.042\\
                & 96741.38 & 6.8 & 1350 & 0.39 & 0.560 & 0.056\\
                & 96744.55 & 7.8 & 110 & 0.41 & 0.049 & 0.006\\
                & 96755.51 & 6.9 & $-$ & $-$ & $<0.004$ &$-$ \\
                & 97582.80 & 3.8 & $-$ & $-$ & $<0.003$ & $-$\\
                & 108893.96 & 5.1 & 170 & 0.39 & 0.070& 0.008\\
                & 143865.80 & 21 & $-$ & $-$ & $<0.011$ & $-$\\
                & 145093.71 & 21 & $-$ & $-$ & $<0.11$ & $-$ \\
                & 145097.37 & 22 & 965 & 0.32 & 0.329 & 0.050\\
                & 145103.15 & 22 & 1090 & 0.31 & 0.360 & 0.054\\
                & 145126.19 & 22 & $-$ & $-$ & $<0.012$ & $-$\\
                & 145126.39 & 22 & $-$ & $-$ & $<0.012$ & $-$\\
                & 145131.85 & 23 & $-$ &  $-$ & $<0.012$ & $-$\\
\hline
L1521E  & 95914.31 & 6.9 & $-$ & $-$ & $< 0.005$ & $-$ \\
                & 96739.36 & 6.7 &  900  & 0.46 & 0.442 & 0.044 \\
                & 96741.38 & 7.5 &  1165  & 0.47 & 0.583 & 0.058 \\
                & 96744.55 & 5.7 &  130  & 0.52 & 0.073 & 0.008 \\
                & 96755.51 & 8.1 & 35   & 0.45 & 0.016 & 0.004 \\
                & 97582.80 & 3.5 & 12 & 0.78 & 0.010 & 0.002 \\
                & 108893.96 & 5.3 &  255   & 0.39 & 0.106 & 0.007 \\
                & 143865.80 & 20 & $-$ & $-$ & $<0.012$ & $-$ \\
                & 145093.71 & 20 & 115  & 0.30 & 0.036 & 0.007 \\
                & 145097.37 & 22 &  960 & 0.36 & 0.367 & 0.055 \\
                & 145103.15 & 18 & 1135  & 0.34 & 0.403 & 0.061\\
                & 145126.19 & 19 &  70 & 0.26 & 0.019 & 0.006 \\
                & 145126.39 & 19 & $-$ & $-$ & $<0.011$ & $-$\\
                & 145131.85 & 20 & $-$ & $-$ & $<0.012$ & $-$ \\
                & 156602.41 & 30 & $-$ & $-$ & $<0.017$ & $-$ \\
                & 157246.06 & 34 & $-$ & $-$ & $< 0.019$ & $-$ \\
                & 157270.84 & 31 & 375   & 0.37 & 0.147 & 0.024\\
                & 157272.37 & 33 & $-$ & $-$ &  $<0.018$ & $-$ \\
                & 157276.06 & 30 &  210  & 0.27 & 0.060 &  0.011 \\
\hline
L1512   & 95914.31 & 5.7 & $-$ & $-$ &  $<0.004$ & $-$ \\
                & 96739.36 & 7.6 &  380  & 0.33 & 0.133 & 0.014 \\
                & 96741.38 &  6.5 & 515  & 0.34 & 0.186 & 0.019 \\
                & 96744.55 & 6.7 &  40  & 0.32 & 0.013 & 0.003 \\
                & 96755.51 & 6.2 & $-$ & $-$ &  $<0.004$ & $-$ \\
                & 97582.80 & 4.3 & $-$ & $-$ &   $<0.003$ & $-$ \\
                & 108893.96 & 4.9 &  70 & 0.33 & 0.026 & 0.003\\
                & 143865.80 & 23 & $-$ & $-$ &  $<0.013$ & $-$ \\
                & 145093.70 & 25 & $-$ & $-$ &  $<0.015$ & $-$ \\
                & 145097.37 & 25 & 485  & 0.20 & 0.103 &  0.016\\
                & 145103.15 & 20 & 510 & 0.22 &  0.119 & 0.018 \\
                & 145126.19 & 22 & $-$ & $-$ &  $<0.013$ & $-$\\
                & 145126.39 & 22 & $-$ & $-$ &  $<0.013$ & $-$ \\
                & 145131.86 & 22 & $-$ & $-$ &  $<0.022$ & $-$ \\
                & 156602.41 & 28 & $-$ & $-$ &  $<0.016$ & $-$ \\
                & 157246.06 & 31 & $-$ & $-$ &  $<0.018$ & $-$ \\
                & 157270.84 & 28 & 165 & 0.30 & 0.053 &  0.011\\
                & 157272.37 & 27 & $-$ & $-$ &  $<0.015$ & $-$ \\
                & 157276.06 & 28 & $-$ & $-$ &  $<0.016$ & $-$ \\
\hline
L1517B    & 95914.31 & 6.6 & $-$ & $-$ &  $< 0.005$ & $-$ \\
                & 96739.36 & 7.4 & 365  & 0.37 & 0.143 &  0.015 \\
                & 96741.38 & 7.4 &  485 & 0.37 & 0.192 &  0.019 \\
                & 96744.55 & 4.7 &  25 & 0.33 & 0.009 &  0.002 \\
                & 96755.51 & 6.1 & $-$ & $-$ &  $<0.004$ & $-$ \\
                & 97582.80 & 4.0 & $-$ & $-$ &  $<0.003$ & $-$ \\
                & 108893.96 & 5.3 & 55 & 0.32 & 0.019 & 0.003 \\
                & 143865.80 & 18 & $-$ & $-$ &  $<0.010$ & $-$ \\
                & 145093.70 & 18 & $-$ & $-$ &  $<0.010$ & $-$ \\
                & 145097.37 & 20 & 355  & 0.23 & 0.087 &  0.014\\
                & 145103.15 & 18 &  405 & 0.25 & 0.108 &  0.017\\
                & 145126.19 & 21 & $-$ & $-$ &  $<0.013$ & $-$  \\
                & 145126.39 & 21 & $-$ & $-$ &  $<0.013$ & $-$  \\
                & 145131.86 & 20 & $-$ & $-$ &  $<0.011$& $-$  \\
                & 156602.41 & 22 & $-$ & $-$ &  $<0.012 $ & $-$ \\
                & 157246.06 & 21 & $-$ & $-$ &  $<0.012$ & $-$ \\
                & 157270.84 & 24 &  160  & 0.24 & 0.041 &  0.008 \\
                & 157272.37 & 23 & $-$ & $-$ & $<0.013$  & $-$  \\
                & 157276.06 & 25 & $-$ & $-$ &  $0.014$ & $-$  \\
\hline
\end{longtable}
\end{longtab}

\begin{table}[h]
\centering
  \caption{\label{coldenshco} HCO column densities  for $T_{\rm ex}$ = 4\,K with 1-$\sigma$ error bars.}
  \begin{tabular}{lcccc}
\hline\hline
Source &  $N$(HCO)   & $N$(HCO) min  & $N$(HCO) max  & $\chi^2_{\rm min}$ \\
              & (cm$^{-2}$)            & (cm$^{-2}$)            & (cm$^{-2}$)  \\
\hline
L1689B  & 1.3 10$^{13}$ & 1.2 10$^{13}$ & 1.4 10$^{13}$ & 0.05\\
L1495A-S & 7.0 10$^{12}$ & 6.4 10$^{12}$ & 7.7 10$^{12}$ & 1.6\\
L429            & 3.0 10$^{12}$ & 2.7 10$^{12}$ & 3.3 10$^{12}$ & 7.0\\
L1709A  & 5.5 10$^{12}$ & 5.0 10$^{12}$ & 5.9 10$^{12}$ & 1.2\\
TMC2    & 2.3 10$^{12}$ & 2.1 10$^{12}$ & 2.6 10$^{12}$         & 8.3 \\
L1517B  & 7.1 10$^{11}$ & 6.1 10$^{11}$ & 8.0 10$^{11}$ & 1.6 \\
L1512   & 9.6 10$^{11}$ & 8.5 10$^{11}$ & 1.1 10$^{12}$ & 1.9 \\
L1521E  & 2.0 10$^{12}$ & 1.7 10$^{12}$ & 2.2 10$^{12}$ & 6.2 \\
\hline
  \end{tabular}
\end{table}

\begin{table}[h]
  \caption{\label{coldensch3o}CH$_3$O (A+E) column densities  for $T_{\rm ex}$ = 8\,K.  Upper limits are 3-$\sigma$.}
\centering
  \begin{tabular}{lcccc}
  \hline\hline
Source &  $N$(CH$_3$O)  & $N$(CH$_3$O) min  & $N$(CH$_3$O) max  & $\chi^2_{\rm min}$ \\
              & (cm$^{-2}$)            & (cm$^{-2}$)            & (cm$^{-2}$)  \\
\hline
L1689B  & 1.2 10$^{12}$ & 1.1 10$^{12}$ & 1.3 10$^{12}$ & 6.2\\
L1495A-S & 8.6 10$^{11}$ & 7.2 10$^{11}$ & 1.0 10$^{12}$ & 0.04\\
L429            & $<$ 3.4 10$^{11}$     & --                            & --                              & --\\
L1709A  & 5.2 10$^{11}$         & 4.3 10$^{11}$ & 6.1 10$^{11}$ & 8.1 \\
TMC2    & 5.6 10$^{11}$ & 3.8 10$^{11}$ & 7.4 10$^{11}$ & 0.2 \\
L1517B  & $<$ 1.9 10$^{11}$ & --                                & --                            & -- \\
L1512   & $<$ 2.0 10$^{11}$ & --                                & --                            & -- \\
L1521E  & $<$ 2.0 10$^{11}$ & --                                & --                            & -- \\
\hline
   \end{tabular}
\end{table}

\begin{table*}[h]
  \caption{\label{coldensch3oh}Best-fit CH$_3$OH (A and E) column densities}
\centering
  \begin{tabular}{lcccccc}
  \hline\hline
Source &  $T_{\rm k}$  &        $n_{\rm H_2}$   & $N$(A-CH$_3$OH)  & $N$(E-CH$_3$OH) & $N$(CH$_3$OH) & reduced $\chi^2_{\rm min}$ \\
              & (K)            & (cm$^{-3}$)            & (cm$^{-2}$) & (cm$^{-2}$) & (cm$^{-2}$)  \\
\hline
L1689B  & 8.5           & 3.6 10$^{5}$          & 6.2 10$^{13}$ & 5.9 10$^{13}$ & 1.2 10$^{14}$   & 46.9\\
L1495A-S & 10            & 5.2 10$^{5}$          & 2.6 10$^{13}$ & 3.5 10$^{13}$ & 6.1 10$^{13}$   & 6.9\\
L429            & 6             & 5.3 10$^{5}$          & 4.7 10$^{13}$ & 3.0 10$^{13}$   & 7.7 10$^{13}$ & 9.3\\
L1709A  & 8             & 3.3 10$^{5}$          & 3.9 10$^{13}$         & 3.2 10$^{13}$   & 7.1 10$^{13}$ & 25.5 \\
TMC2    & 6             & 1.7 10$^{5}$          & 2.2 10$^{13}$ & 1.9 10$^{13}$ & 4.1 10$^{13}$   & 0.2 \\
L1517B    & 5.5          & 1.8 10$^{5}$          & 6.4 10$^{12}$         & 5.0 10$^{12}$        & 1.1 10$^{13}$         & 13.8 \\
L1512   & 5.5           & 5.8 10$^{5}$          & 6.0 10$^{12}$ & 4.8 10$^{12}$ & 1.1 10$^{13}$   & 5.8 \\
L1521E  & 6             & 4.8 10$^{5}$          & 2.2 10$^{13}$ & 1.8 10$^{13}$ & 4.0 10$^{13}$   & 19.9 \\
  \hline
   \end{tabular}
\end{table*}

\begin{table*}[h]
  \caption{\label{1sigmach3oh}1-$\sigma$ uncertainties for CH$_3$OH (A and E) column densities}
\centering
  \begin{tabular}{lcccccc}
  \hline\hline
Source &  $T_{\rm k}$  &        $n_{\rm H_2}$   & $N$(A-CH$_3$OH)  & $N$(E-CH$_3$OH) & $N$(CH$_3$OH)  \\
              & (K)            & (cm$^{-3}$)            & (cm$^{-2}$) & (cm$^{-2}$) & (cm$^{-2}$)  \\
\hline
L1689B  & 8 -- 9                & 3.1 -- 4.1 10$^{5}$           & 5.5 -- 7.4 10$^{13}$   & 5.1 -- 7.4 10$^{13}$  & 1.1 -- 1.5 10$^{14}$ \\
L1495A-S & 8 -- 13               & 3.8 -- 8.2 10$^{5}$           & 2.2 -- 3.3  10$^{13}$  & 2.9 -- 4.5 10$^{13}$  & 5.2 -- 7.8 10$^{13}$ \\
L429            & 5.5 -- 7.5    & 3.0 -- 9.6 10$^{5}$           & 2.4 -- 20.0 10$^{13}$  & 2.1 -- 4.5 10$^{13}$  & 4.6 -- 24 10$^{13}$ \\
L1709A  & 7 -- 10.5             & 2.2 -- 5.0 10$^{5}$           & 2.9 -- 5.4 10$^{13}$   & 2.4 -- 4.3 10$^{13}$  & 5.4 -- 9.5 10$^{13}$ \\
TMC2    & 4.5 -- 8.5    & 0.3 -- 4.9 10$^{5}$           & 1.3 -- 17 10$^{13}$   & 1.2 -- 5.2 10$^{13}$    & 2.5 -- 21 10$^{13}$ \\
L1517B  & 4 -- 14               & 0.1 -- 7.3 10$^{5}$           & 4.0 -- 15 10$^{12}$    & 3.6 -- 9.3 10$^{12}$  & 0.8 -- 2.4 10$^{13}$ \\
L1512   & 4 -- 7                & 2.0 -- 27 10$^{5}$            & 4.0 -- 15 10$^{12}$    & 3.5 -- 9.5 10$^{12}$  & 0.8 -- 2.4 10$^{13}$ \\
L1521E  & 4.5 -- 8              & 3.2 -- 7.9 10$^{5}$           & 1.4 -- 5.0 10$^{13}$   & 1.3 -- 3.5 10$^{13}$  & 2.8 -- 8.5 10$^{13}$ \\
 \hline
   \end{tabular}
\end{table*}

\section{Discussion}

\label{discus}
In each source, the linewidths of the different species are very similar so that they are most probably coexistent. The abundances of H$_2$CO, CH$_3$O and CH$_3$OH with respect to HCO are reported in Table\,\ref{abunratio}. In the table we chose HCO as the reference because it was observed and detected in all sources. The H$_2$CO data come from the literature, but only half of the sources in our sample have been observed in this species. The H$_2$CO column densities for L1689B, L429, and L1709A are 1.3 10$^{14}$\,cm$^{-2}$, 2.8 10$^{13}$\,cm$^{-2}$, and 5.6 10$^{13}$\,cm$^{-2}$, respectively \citep{Bacmann:2003p1218}. For L1517B, the column density was integrated from the H$_2$CO radial abundance profile described in \citet{Tafalla:2006p3988}, and found to be 5.2 10$^{12}$\,cm$^{-2}$.

While the column densities in the species vary by over an order of magnitude between the different sources, their ratios show relatively little variations. Indeed, we find that within a factor of 2, the HCO:H$_2$CO:CH$_3$O:CH$_3$OH ratios are close to 10:100:1:100. For H$_2$CO, the comparison between  L1689B, L429, and L1709A, on the one hand, and L1517B, on the other, should be made cautiously since the column densities have been determined with different methods. Besides this, the lines observed in L1517B have very high optical depths and are self-absorbed, and no optically thin isotope was observed. This H$_2$CO column density value could therefore be underestimated. The result that HCO is about ten times more abundant than CH$_3$O seems robust\footnote{\citet{Cernicharo:2012p4397} find [HCO]/[CH$_3$O]$\,\approx 4$ in B1-b, but they have assumed an excitation temperature of 10\,K for HCO, which leads to lower column densities than the excitation temperature of 4\,K that we have derived.}, and the upper limits derived in the sources where  CH$_3$O was not detected are also consistent with this number, although they may not be stringent enough. TMC\,2 has a higher CH$_3$O relative abundance than the other sources, but its detection is marginal since the strongest lines are only at the 3-$\sigma$ level. Similarly, CH$_3$OH is 10 to 25 times more abundant than HCO  and the H$_2$CO abundances are of the same order of magnitude as in CH$_3$OH. The latter is consistent with previous results obtained by \citet{Guzman:2013p4399}, who found a H$_2$CO/CH$_3$OH ratio of 0.9 at the dense core position in the Horsehead nebula. 

Available chemical models have difficulty accounting for our observational results. The model recently proposed by \citet{Vasyunin:2013p4398} shows the abundances of complex organic molecules,  as well as CH$_3$O, CH$_3$OH, and H$_2$CO. The abundances of H$_2$CO and CH$_3$OH are similar only towards steady state, at which times  
the CH$_3$OH/CH$_3$O abundance ratio is $\sim$\,30,  slightly lower (a factor of $\sim$\,3)  than our observations. 
In the model, CH$_3$O is over an order of magnitude more abundant than molecules like CH$_3$CHO or CH$_3$OCHO, whereas in L1689B CH$_3$O is observed to be an order of magnitude less abundant than these species \citep{Bacmann:2012p4015}.  In the model of \citet{Balucani:2015p5368}, the CH$_3$OH/CH$_3$O abundance ratio is close to our observed values for ages of at most a few 10$^5$ years. This is unlikely to be the age of all cores in our sample, as the lifetime of cores is estimated to be around 10$^6$ years \citep{2014prpl.conf...27A}. At average core ages, the model predicts no detectable amount of methanol in the gas phase.

In the following sections, we explore various chemical routes to explain the observed HCO:H$_2$CO:CH$_3$O:CH$_3$OH abundance ratios, focussing on the dominant formation and destruction reactions. Square brackets refer to molecular abundances with respect to H$_2$.

\begin{table}[h]
  \caption{\label{abunratio}Abundance ratios with respect to HCO. A dash appears when there is no available measurement. Upper limits are 3-$\sigma$.}
\centering
  \begin{tabular}{lccc}
  \hline\hline
  Source                & H$_2$CO/HCO   & CH$_3$O/HCO   & CH$_3$OH/HCO \\
  \hline
  L1689B                &       10                      &       0.09                    &       9                       \\
  L1495A-S      &       --                      &       0.12                    &       9                       \\
  L429          &       9                       &       $< 0.11$                &         26                      \\
  L1709A                &       10                      &       0.09                    &       13                      \\
  TMC\,2                &       --                      &       0.24                    &       18                      \\
  L1517B                &       7                       &       $< 0.27$                &       15                      \\
  L1512         &       --                      &       $< 0.21$                &       11                      \\
  L1521E                &       --                      &       $<0.10$         &       20                      \\
  \hline
  \end{tabular}
\end{table}

\subsection{Grain surface formation route}

According to current understanding
\citep{Geppert:2006p4196,Garrod:2006p4851}, methanol forms on grain
surfaces from the successive additions of H atoms to CO, following
CO\,$\rightarrow$\,HCO\,$\rightarrow$\,H$_2$CO\,$\rightarrow$\,CH$_3$O\,$\rightarrow$\,CH$_3$OH.
This sequence of reactions has been studied in the laboratory at 10\,K
by \citet{Watanabe:2002p3383} and both at 10\,K and at 3\,K by
\citet{Pirim:2010p4906} and \citet{Pirim:2011p4136}. The intermediate radicals HCO
and CH$_3$O are not observed in the experiments by
\citet{Watanabe:2002p3383}, but they are detected at 10~K in those by
Pirim et al., who co-deposited CO and H in their study, whereas
\citet{Watanabe:2002p3383} injected H only after CO had been already
deposited. \citet{Pirim:2011p4136} have also shown that in the
presence of water, the HCO conversion into H$_2$CO and that of H$_2$CO
into CH$_3$O and CH$_3$OH are enhanced. Water thus acts as a catalyst
of CO hydrogenation. As a result, in interstellar grain mantles
dominated by water, only the most stable intermediates H$_2$CO and
CH$_3$OH are expected to be present with significant abundances.

In prestellar cores, at 10\,K, the successful grain-surface hydrogenation of CO into H$_2$CO and CH$_3$OH will depend on the competition between H addition on the grains and 
CO  desorption: if CO can desorb before H accretes on the grain and react with CO, no CO hydrogenation will take place. The same applies to the other H$_2$CO and CH$_3$OH precursors, such as HCO and CH$_3$O. The timescale for hydrogenation on a grain surface in the dense interstellar medium is very short, since H atoms physisorbed on the grains can scan the whole surface many times before evaporating back into the gas phase \citep{Tielens:1982p3002}, and it is therefore not a limiting factor for the hydrogenation of the species on the grains. And typically,  one H atom will arrive on a grain every day. Timescales for desorption for the methanol precursors either by cosmic ray impacts or by secondary UV photons are of the order of 10$^4-10^5$ years \citep{Tielens:2005p4957,Banerji:2009p4962}. 
This means that the radicals HCO and CH$_3$O will have ample time to be hydrogenated before they can be desorbed.

The hydrogenation of the unstable radicals HCO and CH$_3$O into H$_2$CO and CH$_3$OH, respectively, is believed to take place quickly, whereas the formation of these radicals from CO and H$_2$CO has a barrier \citep[about 2500\,K and 2200\,K, respectively $-$][]{Garrod:2008p2520,Osamura:2005p4965}, according to both theoretical calculations and experiments \citep{Peters:2013p4920,Wang:1973p4921}\footnote{Recent experiments by \citet{Pirim:2011p4905} show, however, that H+CO proceeds with no or a very small barrier at 3\,K, so that the issue is not completely settled.}. It is therefore likely that H atoms arriving on the grain surface will usually react with HCO or CH$_3$O rather than with CO or H$_2$CO, leading to the most stable products.

Taking all these arguments into account, it is therefore possible that  the amounts of HCO and CH$_3$O  on interstellar grain surfaces are very small and that the HCO and CH$_3$O seen in the gas phase is the product of gas-phase processes. We present below possible gas-phase formation routes for HCO and CH$_3$O and discuss their validity with respect to our observational results.
In the simple chemical modelling presented below, we re-evaluated the reaction rates for charge-dipole reactions using the "locked dipole" approximation. The method is described in Appendix \ref{capturerates}. 
The adopted reaction rates are listed in Table\,\ref{reactionrates}.

\subsection{Neutral-neutral pathways}

Recent experimental results by \citet{Shannon:2013p4217} show that neutral-neutral reactions such as CH$_3$OH + OH $\rightarrow$ CH$_3$O + H$_2$O accelerate at low temperatures ($T=63$\,K in their experiment) despite having a barrier. Those results were confirmed and extended to lower temperatures by \citet{gomezmartin:2014p4918}. Similarly, other neutral-neutral reactions, such as the barrierless reaction CH$_3$OH + CN \citep{Sayah:1988p4966}, also produce CH$_3$O. HCO can be formed with the corresponding reactions involving H$_2$CO instead of CH$_3$OH and also with CH$_3$OH + C \citep{Shannon:2014aa}. The following reaction may therefore be considered for the formation of CH$_3$O:\\
\begin{equation}
\mathrm{CH_3OH+ OH \xrightarrow{k_1} CH_3O + H_2O}
\label{ch3ohoh}
\end{equation}
and for its destruction
\begin{equation}
\mathrm{CH_3O + HX^+ \xrightarrow{k_2} CH_3OH^+ + X}
\label{ch3odestruc}
\end{equation}
where HX$^+$ stands in fact for any major positive ion which can easily give a proton (e.g. H$_3^+$). This reaction has so far not been considered in any of the public astrochemical networks \citep[KIDA, UMIST $-$][]{McElroyD:2013ki}. At steady state, the abundance of CH$_3$O is given by
\begin{equation}
\mathrm{[CH_3O]=\frac{k_1[CH_3OH][OH]}{k_2 [HX^+]}}
\end{equation}
where the square brackets indicate abundances (with respect to H$_2$).
Assuming a typical abundance of 10$^{-8}$ for positive ions, consistent with the ionisation fraction in dark cores \citep{Flower:2005p1891,Flower:2006p1295,Pagani:2009p1030}, an average abundance of 8\,10$^{-8}$ for OH \citep{Troland:2008p1151,Crutcher:1979p4968},  we find that [CH$_3$O]$=8\,$k$_1$/k$_2$[CH$_3$OH]. 
With the values of the reaction rates from Table\,\ref{reactionrates}, this leads to an abundance of [CH$_3$O$]\,\sim\,0.007\,$[CH$_3$OH], which is slightly lower than but close to what is measured observationally ([CH$_3$O]\,$=1/100$\,[CH$_3$OH]). In this simple scheme, we also assumed that CH$_3$O is the sole product of reaction (\ref{ch3ohoh}), as explained in \citet{Shannon:2013p4217}. The OH abundance given here is, however, uncertain as it is the typical abundance measured from low critical density lines over a large beam (3 arcminutes for the Arecibo telescope at 1.6\,GHz) and may not be representative of what is seen on smaller scales towards the centres of prestellar cores.
Other destruction routes, such as the depletion of CH$_3$O on grains, are not efficient compared with ion-molecule reactions like those mentioned above. Indeed, depletion rates  onto grains are about $\rm k_{\rm depl} = 3\,10^{-6}$\,cm$^{3}$s$^{-1}$ for species with a similar mass to CO \citep{Hasegawa:1993tv}, and grain abundances [g] of the order of  2 10$^{-12}$ \citep{Li:2003jz}.  The disappearance of CH$_3$O due to depletion is therefore less efficient than its destruction by proton donors by a factor:

\begin{equation*}
\frac{{\rm k}_2\, [\mathrm{HX}^+]}{\mathrm{k_ {\rm depl}} \,[{\rm g}]} \approx 100.
\end{equation*}

For HCO formation, the corresponding neutral-neutral route is\\
\begin{equation}
\mathrm{H_2CO+ OH \xrightarrow{k^{\prime}_1} HCO + H_2O}
\label{h2cooh}
\end{equation}
and for its destruction
\begin{equation}
\mathrm{HCO + HX^+ \xrightarrow{k^{\prime}_2} HCOH^+ + X}
\label{hcodestruc}
.\end{equation}

The abundance of HCO is therefore [HCO]=8\,k$^\prime_1$/k$^\prime_2$ [H$_2$CO]. No measurement of the rate coefficient for reaction (\ref{h2cooh}) exists at low temperatures, however measurements at 230\,K give k$^{\prime}_1=10^{-11}$\,cm$^{3}$s$^{-1}$ %
\citep{Baulch:2005p5096}. Using this value, we derive [HCO]\,$=1.6\,10^{-3}$\,[H$_2$CO]. 
This is about two orders of magnitude less than the value we measure. Again, we have assumed that HCO is the sole product of reaction (\ref{h2cooh}), as suggested in \citet{Baulch:2005p5096}.  

In our simple model, the abundance ratio [HCO]/[CH$_3$O] is equal to (k$^\prime_1$/k$_1$)\;(k$_2$/k$^\prime_2$)\;[H$_2$CO]/[CH$_3$OH]. This is independent of any assumption on the OH or proton donor abundance. Compared with our observed ratio, this means that k$^\prime_1$ would have to be 4\,10$^{-10}$\,cm$^3$s$^{-1}$  at 10\,K in order to explain our results. This value is close to the upper limit for neutral-neutral reactions, but it cannot be ruled out, since similarly high reaction rate values between neutrals have been reported at low temperatures in, for example, \citet{Sims:1994p5303}.

\subsection{Ion-molecule fomation route}

Ion-molecule reactions represent an alternative, faster pathway to form CH$_3$O and HCO. For such a scheme, a possible formation route to CH$_3$O would be\\
\begin{equation}
\mathrm{CH_3OH+ HX^+ \xrightarrow{\mathit{r}\;k_3} CH_3OH_2^+ + X}
\label{ch3ohion}
\end{equation}
followed by the dissociative recombination,
\begin{equation}
\mathrm{CH_3OH_2^+ + e^- \xrightarrow{\mathit{f}\;k_4} CH_3O + H_2}
\label{drch3oh}
,\end{equation}
where, again, HX$^+$ stands for a positive ionic proton donor, like H$_3^+$.  
In the present reaction rate databases, reaction (\ref{drch3oh}) was omitted as a production route for CH$_3$O.
The destruction route for CH$_3$O is the same as in the previous section.

We denote k$_3$ and k$_4$ as the destruction rate of the reactants in reactions\,(\ref{ch3ohion}) and (\ref{drch3oh}), and $r$ and $f$ the branching ratios for the products CH$_3$OH$_2^+$ and CH$_3$O, respectively.

At steady state, the CH$_3$O abundance can be written as
\begin{equation}
\mathrm{[CH_3O]=\frac{\mathit{f}\;k_4 [CH_3OH_2^+][e^-]}{k_2 [HX^+]}}
\label{ch3oabund}
.\end{equation}

The CH$_3$OH$_2^+$ abundance is
\begin{equation}
\mathrm{[CH_3OH_2^+]=\frac{\mathit{r}\;k_3 [CH_3OH][XH^+]}{k_4 [e^-]}}
\label{ch3oh2abund}
.\end{equation}
Substituting (\ref{ch3oh2abund}) in (\ref{ch3oabund}), we find that, at steady-state,
\begin{equation}
\mathrm{[CH_3O]=\frac{\mathit{f\;r}\;k_3}{k_2}[CH_3OH]}
.\end{equation}

According to \citet{Geppert:2006p4196}, the branching ratio $f$ for CH$_3$O production from CH$_3$OH$_2^+$ dissociative recombination is about 6\%. The value for $r$ derived from the KIDA\footnote{http://kida.obs.u-bordeaux1.fr} database \citep{Wakelam:2012p4970} is 25\% (assuming HX$^+$ to be mostly H$_3^+$).  
This yields
\begin{equation*}
\mathrm{[CH_3O] = 1.2\;10^{-2} [CH_3OH].}
\end{equation*}
This can account for the CH$_3$O observed in our sources.  
Following equation (\ref{ch3oh2abund}), the abundance of CH$_3$OH$_2^+$  is
\begin{equation*}
\mathrm{[CH_3OH_2^+] = 1.5\,10^{-3} [CH_3OH] }
\end{equation*}
where we have 
made the approximation that [HX$^+$]$\approx$[e$^-$]. \\

Similar reactions can be considered for the formation of HCO:\\
\begin{equation}
\mathrm{H_2CO + XH^+ \xrightarrow{k^\prime_3} H_2COH^+ + X}
\end{equation}
\begin{equation}
\mathrm{H_2COH^+ + e^- \xrightarrow{\mathit{f^\prime} k^\prime_4} HCO + H + H}
,\end{equation}
and again, reaction (\ref{hcodestruc}) accounts for the destruction of HCO. As before, k$^{\prime}_4$ is the rate at which H$_2$COH$^+$ is destroyed by electrons and $f^\prime$ the branching ratio towards the HCO product. Unlike methanol, the reaction of H$_2$CO with a proton donor such as H$^+_3$ gives H$_2$COH$^+$ as the only product, according to the KIDA database. The abundance of HCO can be written as
\begin{equation}
\mathrm{[HCO]=\frac{\mathit{f^\prime}\;k^\prime_4 [H_2COH^+][e^-]}{k^\prime_2[HX^+]}}
\label{hcoabund}
.\end{equation}
In turn, the H$_2$COH$^+$ abundance  is
\begin{equation}
\mathrm{[H_2COH^+] = \frac{k^\prime_3 [H_2CO][XH^+]}{k^\prime_4 [e^-]}}\label{h2cohabund}
,\end{equation}
and finally, substituting eq. (\ref{h2cohabund}) into eq. (\ref{hcoabund}) the HCO abundance is
 \begin{equation}
\mathrm{[HCO]=\frac{\mathit{f^\prime}\;k^\prime_3}{k^\prime_2} [H_2CO]}.\end{equation}
\citet{Hamberg:2007p4982} carried out experimental measurements of the dissociative recombination of protonated formaldehyde H$_2$COH$^+$. They found that fragment products where the C$-$O bond is preserved make up 92\% of the total yield of the reaction. However, they were not able to distinguish between the various products containing a C$-$O bond (HCO, CO, or H$_2$CO), so that the branching ratio they give is only an upper limit for $f^\prime$. We assume $f^\prime$=10\% for the formation of HCO from the dissociative recombination of  H$_2$COH$^+$.

With this hypothesis and the rate coefficient values from Table\,\ref{reactionrates}, the HCO abundance is\\

\begin{equation}
\mathrm{[HCO]=0.14\;[H_2CO]}
,\end{equation}
which is close to what is observed in our sources. A gas-phase route with ion-molecule reactions could therefore produce both enough HCO and CH$_3$O  to account for the observed abundances. In what precedes, we have neglected that protonated formaldehyde can additionally be formed from the reaction of methanol with H$_3^+$; however, given the rate for that reaction ($\sim 6\, 10^{-9}$\,cm$^3$s$^{-1}$), this would increase the abundance by only 50\%. The abundance of protonated formaldehyde is then
\begin{equation}
\mathrm{[H_2COH^+]=7\,10^{-3}[H_2CO]}
\end{equation}
where we have taken k$^\prime_4=9.9\,10^{-6}$\,cm$^3$s$^{-1}$, the value derived at 10\,K from \citet{Hamberg:2007p4982}.

The extreme simplicity of our chemical model, as well as the uncertainties on the reaction rates (most of which have been calculated at 10\,K and not measured) or branching ratios implies that the abundances we derive should be taken as rough estimates and not as precise values. Moreover, the steady-state hypothesis we have made might not be verified in prestellar cores, where dynamical and chemical timescales are of the same order of magnitude.  

Time-dependent models with an elaborated network will be investigated in a future, dedicated study. Still, an interesting test of this ion-molecule scheme would come from the detection and abundance measurements of CH$_3$OH$_2^+$ and H$_2$COH$^+$. The abundances of these species derived from our model are 1.5\,10$^{-3}$\,[CH$_3$OH] and 7\,10$^{-3}$\,[H$_2$CO], respectively, which would be 1.5\,10$^{-12}$ and 7\,10$^{-12}$ for a source like L1689B. CH$_3$OH$_2^+$ has never been detected, since its rotational frequencies have not been determined, but the abundance  of H$_2$COH$^+$ was determined by \citet{Ohishi:1996p4278} in various types of sources and found to be around 10$^{-11}$ in some of them. Though H$_2$COH$^+$ was looked for in dark clouds, it was not detected by these authors, and because no upper limits were given in the article, it is not possible to say whether these non-detections invalidate our model. \\

Although neutral-neutral reactions are generally slower than proton transfer in the gas phase, they can be competitive if the reactant is very abundant. Therefore, a possible alternative destruction route for HCO and CH$_3$O is the reaction of these radicals with H atoms, following\\
\begin{eqnarray}
\mathrm{ CH_3O  +  H \xrightarrow{k_5} H_2CO  + H_2}\label{ch3odesth}\\
\mathrm{ HCO  +  H \xrightarrow{k^\prime_5}  CO +  H_2}\label{hcodesth}
.\end{eqnarray}

The only measurements of the rate constants for these reactions are at 300\,K and above \citep{Baulch:2005p5096}, and  it is extremely uncertain to extrapolate them at 10\,K, since they could differ by an order of magnitude or more. Bearing this in mind, we try to estimate in the following the effect of these reactions. The corresponding abundances of CH$_3$O and HCO are given by\\
\begin{eqnarray}
\mathrm{[CH_3O]=\frac{\mathit{f\;r}\; k_3 [CH_3OH][HX^+]}{k_5 [H]}}\\
\mathrm{[HCO]=\frac{\mathit{f^\prime} k^\prime_3 [H_2CO][HX^+]}{k^\prime_5 [H]}}
.\end{eqnarray}
The atomic H density is typically around $n_{\rm H}=2-6\,$cm$^{-3}$
independent of the density \citep{Goldsmith:2005p5113}, and the atomic
H abundance ([H]=$n_{\rm H}/n_{\rm H_2}$) is therefore between a few
10$^{-6}$ and a few 10$^{-5}$. If the values of the rate constants at
300\,K hold at 10\,K (see Table\,\ref{reactionrates}), this would lead
to $\mathrm{[CH_3O]=2.3\,10^{-7} [CH_3OH]/[H]}$ and
$\mathrm{[HCO]=4.7\,10^{-7} [H_2CO]/[H]}$, which is a ratio
[HCO]/[CH$_3$O]\,=\,2\,[H$_2$CO]/[CH$_3$OH]\,$\sim$\,2. This is not
consistent with our observations. We note, however, that increasing
$\rm{k_5/k^\prime_5}$ by a factor of 5 would produce the observed ratio.

Similar destruction reactions with other radicals like atomic oxygen are also possible with
similar reaction rates. The abundance of O in prestellar cores is extremely uncertain, but
it is expected 
drop below 10$^{-6}$ after 2\,10$^5$ years \citep{Aikawa:2003p1176} so that the main neutral-neutral destruction route of HCO and CH$_3$O 
should be with H instead. If O were as abundant as H, the destruction rate of CH$_3$O and HCO by neutrals (assuming these rates are similar
in the case of O and in the case of H) would by a factor of two higher with respect to a destruction by H alone, but this would not affect our conclusions.


\begin{table*}
\caption{List of reactions and adopted rates at 10\,K and adopted branching ratios for reactions with several output channels.} 
\label{reactionrates}
\begin{tabular}{llll}
\hline\hline
Reaction & Rate coefficient (cm$^3$s$^{-1}$) & Branching ratio & Reference\\
\hline
$\mathrm{CH_3OH + OH \rightarrow CH_3O + H_2O}$ & k$_1$ = 5\;10$^{-11}$ & &  rate at 50\,K, \citet{gomezmartin:2014p4918}\\
$\mathrm{CH_3OH + H_3^+ \rightarrow CH_3OH_2^+ + H_2}$ & k$_3$ = 5\;10$^{-8}$ & $r$ = 0.25 & capture rate at 10\,K (Appendix \ref{capturerates})\\
$\mathrm{CH_3OH_2^+ + e^- \rightarrow CH_3O + H_2}$ & k$_4$ = 6.7\;10$^{-6}$ & $f$ = 0.06 & rate at 10\,K, \citet{Geppert:2006p4196} \\
$\mathrm{CH_3O + H_3^+ \rightarrow products}$ & k$_2$ = 6\;10$^{-8}$ & & capture rate at 10\,K (Appendix \ref{capturerates})\\

$\mathrm{CH_3O + H \rightarrow products}$ & k$_5$ = 3.3\;10$^{-11}$ &  & rate at 300\,K, \citet{Baulch:2005p5096} \\
$\mathrm{H_2CO + OH \rightarrow HCO + H_2O}$ & k$^\prime_1$ = 10$^{-11}$ & & rate at 230\,K, \citet{Baulch:2005p5096}\\
$\mathrm{H_2CO + H_3^+ \rightarrow H_2COH^+ + H_2}$ & k$^\prime_3$ = 7\;10$^{-8}$ & & capture rate at 10\,K (Appendix \ref{capturerates})\\
$\mathrm{H_2COH^+ + e^- \rightarrow HCO + H_2}$ & k$^\prime_4$ = 9.9\;10$^{-6}$ & $f^\prime$ = 0.10 & rate at 10\,K, \citet{Hamberg:2007p4982}\\
$\mathrm{HCO + H_3^+ \rightarrow products}$ & k$^\prime_2$ = 5\;10$^{-8}$ & & capture rate at 10\,K (Appendix \ref{capturerates})\\
$\mathrm{HCO + H \rightarrow products}$ & k$^\prime_5$ = 1.5\;10$^{-10}$ & & rate at 300\,K, \citet{Baulch:2005p5096}\\
\hline
\end{tabular}
\tablefoot{For reactions with several product channels, only the most relevant channel is given in the table, as well as the value of the branching ratio. The reaction rates listed should be understood as the rates of disappearance of the reactants. The value of the branching ratio $f^\prime$ is uncertain, because the experimental measurement by \citet{Hamberg:2007p4982} only constrains it to be below 90\%.}
\end{table*}

\subsection{Isomerisation of CH$_3$O}

The methoxy radical CH$_3$O has a structural isomer, the hydroxy methyl radical CH$_2$OH, which is more stable by about 42 kJ/mol \citep{Wang:2012dja}. No microwave spectroscopic data are available in the literature, and as a consequence, the rotational transitions of CH$_2$OH have never been detected in the interstellar medium. \citet{Shannon:2013p4217} conclude from their experimental study that CH$_3$O is the dominant product  of reaction (\ref{ch3ohoh}) at low temperatures. The dissociative recombination of CH$_3$OH$_2^+$, on the other hand, leads a priori to both isomers, and the proportion of CH$_3$O with respect to CH$_2$OH has not been determined experimentally. If this reaction should lead to CH$_2$OH in non-negligible amounts, the value we took for the branching ratio $f$ would be overestimated.

Theoretical calculations \citep{Wang:2012dja} predict that CH$_3$O
should isomerise into CH$_2$OH. If this happens on timescales shorter than
chemical timescales in prestellar cores, the isomerisation would
become a major loss channel for CH$_3$O. Unfortunately, experimental
measurements have yielded only non stringent lower limits constraints
on the isomerisation timescales \citep[e.g.][]{Gutman:1982gv} of the order
of 1s at room temperature, and indeed no evidence of isomerisation has
been reported from experimental source.

\subsection{Protostars}

The ion-molecule gas-phase reactions we have considered were applied to cold medium conditions: values of the rate coefficients were taken at 10\,K, ionisation fraction and dominant ions are those found in prestellar cores. Available observations of the young Class 0 protostar \object{IRAS16293-2422} from the TIMASSS survey \citep{Caux:2011p2519} show the presence of HCO most probably originating in much warmer gas than 10\,K \citep[as shown by the HCO linewidths of 2-3 km/s, compared to those of NH seen in the cold envelope $-$][]{Bacmann:2010p2163} but CH$_3$O remains undetected. The integrated line intensities for HCO in this source are given in Table\,\ref{tablehcoproto}. A similar LTE analysis to what is described for the prestellar cores gives an HCO column density of $N(\mathrm{HCO})=1.4\,10^{13}$\,cm$^{-2}$ for $T_{\rm ex}=40$\,K and $N(\mathrm{HCO})=3.6\,10^{13}$\,cm$^{-2}$ for $T_{\rm ex}=80$\,K. According to \citet{Loinard:2000p4986}, the column density of H$_2$CO in that source is $N(\mathrm{H_2CO})=3.4\,10^{14}$\,cm$^{-2}$. These measured abundances therefore also verify the [H$_2$CO]/[HCO] ratio derived from our ion-neutral model described above. The methanol column density in this source is $N(\mathrm{CH_3OH}) = 10^{16}$\,cm$^{-2}$ \citep{Parise:2004p1297}. Assuming $[\mathrm{CH_3O}] = 0.01\,[\mathrm{CH_3OH}]$ gives a column density of $N(\mathrm{CH_3O})=10^{14}$\,cm$^{-2}$, and with this assumption, CH$_3$O lines are expected to be above 100\,mK, much higher than the rms of the TIMASSS spectral survey (6\,mK, 13\,mK, and 9\, mK at the transition frequencies of methoxy 82.4\,GHz, 137.4\,GHz, and 247.4\,GHz, respectively), independently of the assumed excitation temperature (between 10 \,K and 80\,K). Clearly, the methoxy abundance in this source has to be much lower. Because of the uncertainty on the excitation temperature, it is difficult to give accurate upper limits for the CH$_3$O abundance. For $T_{\rm ex}=40$\,K, methoxy should be detected provided the E-CH$_3$O column density is higher than typically $2\,10^{12}$\,cm$^{-2}$, and for $T_{\rm ex}=80$\,K if $N(\mathrm{E-CH_3O})$ is greater than $5\,10^{12}$\,cm$^{-2}$. 

A possibility of reconciling our model with the non-detection of methoxy in IRAS16293-2422 is to suppose that H$_2$CO and CH$_3$OH and their daughter molecules HCO and CH$_3$O are formed in the prestellar phase. In the hot corino, methanol desorbs from the grains (increasing its column density by one to two orders of magnitude), but the gas-phase chemistry is limited by the short lifetime of the hot core phase, so that the HCO and CH$_3$O column densities are still similar to what is measured in the prestellar phase. Indeed, the timescale for the H$_3^+$ chemistry is typically $t_{\rm chem}=$\,1/$r$k$_3 n(\rm{H_3^+})$, where $n(\rm{H_3^+})$ is the H$_3^+$ density. A typical value for k$_3$ is a few 10$^{-9}$\,cm$^{-3}$s$^{-1}$ at 100\,K  
for $\mathrm{CH_3OH + H_3^+}$. If we take $n(\rm{H_3^+})=10^{-2}$\,cm$^{-3}$, we find a timescale for ion-molecule chemistry with H$_3^+$ of   $t_{\rm chem}\sim$ a few $10^3$\,years. This is shorter than the estimated lifetime of class 0 protostars in the $\rho$ Ophiuchi region of about 45~000 years \citep{Evans:2009p4988}, but larger than the estimated age of the IRAS16293-2422 hot corino \citep[a few hundred years according to][]{Schoier:2002p4241}. These numbers should, however, be taken with caution because age estimates for hot corinos depend on the model and have high uncertainties. Incidently, the HCO column density we measure in the prestellar core L1689B is similar to the one derived in IRAS16293-2422. A CH$_3$O column density of $\sim 10^{12}$\,cm$^{-2}$ as in L1689B would remain undetected with the actual sensitivity of the TIMASSS survey in IRAS16293-2422.  

It is, however, difficult to be conclusive from the observations of only one source. Clearly more observations of HCO and CH$_3$O in the hot cores of low-mass protostars are needed to constrain the chemistry of these species in the warm/hot gas.

\begin{table*}[H]
\caption{\label{tablehcoproto}Line parameters for HCO  observations in the protostellar source IRAS16293-2422. The line intensity $T_{\rm mb}$, the linewidth $\Delta v$, the integrated intensity were determined from Gaussian fits to the line. The error is the quadratic sum of the statistical fit error and a calibration error taken as 10\% at 3\,mm.}
\begin{tabular}{lcccccc}
\hline\hline
Source & Frequency  & rms &  $T_{\rm mb}$  & $\Delta v$ & Integrated intensity& Error on int. intensity \\
               & MHz           & mK &     K                     &   km s$^{-1}$ & K km s$^{-1}$ & K km s$^{-1}$\\
\hline
HCO             & 86670.76 & 6.2 &  37 & 3.0 & 0.117 &  0.021 \\
                & 86708.36 & 6.9 &  38 & 2.0 & 0.081 &  0.017\\
                & 86777.46 & 5.9 & $-$   & $-$ & $<0.029$ & $-$\\
                & 86805.78 & 5.9 & 33 & 1.7 & 0.059  & 0.017 \\
 \hline
\end{tabular}
\end{table*}

\section{Conclusions}

\label{cl}

We observed the species HCO, CH$_3$O, and CH$_3$OH in a sample of
eight prestellar cores and considered H$_2$CO data from the
literature. These species are believed to form on the grain surfaces
by successive hydrogenations of CO. In addition, HCO and CH$_3$O may
play an important role as COM precursors on grain surfaces. The formyl
radical HCO is detected in all sources and the methoxy radical CH$_3$O
in four of them. Since CH$_3$O transitions are very weak and hard to
detect in prestellar cores, it is likely that CH$_3$O is as ubiquitous
as HCO in prestellar cores but that more sensitive observations would
have been needed to detect it in all sources. The column densities of
the observed species vary by an order of magnitude across the source
sample but their abundance ratios with respect to a common reference
like CH$_3$OH are remarkably similar in all the sources. We find that
the abundance ratios HCO\,:\,H$_2$CO\,:\,CH$_3$O\,:\,CH$_3$OH are
close to 10\;:\;100\;:\;1\;:\;100. 

Though it is not possible to
completely exclude the possibility that these ratios can be produced
by successive hydrogenations on grain surfaces, the timescales
involved for grain surface accretion and hydrogenation vs. those for
desorption do not favour a surface chemistry origin for HCO and CH$_3$O. Most
probably, in a mixture of CO and water ice, as is probably the case in
these objects, the main product of the CO hydrogenation reaction is
H$_2$CO and CH$_3$OH and very little HCO or CH$_3$O remain on the
grains at any given time. We examined possible gas-phase formation
routes for HCO and CH$_3$O, thereby introducing two new reactions involving
CH$_3$O that  were omitted in previous models. Neutral-neutral reactions between CH$_3$OH
and radicals like OH have been shown experimentally to form CH$_3$O
and to be efficient at low temperatures. The CH$_3$O abundance
obtained by this process indeed reproduces the observed
[CH$_3$O]/[CH$_3$OH] ratio. On the other hand, the corresponding reaction between H$_2$CO and a radical like OH needs to be even faster, with a
rate exceeding $\sim 4\times 10^{-10}$\,cm$^3$s$^{-1}$ at 10\,K, in order to
account for the observed [HCO]/[H$_2$CO] abundance
ratios. 

Ion-molecule offer a more viable alternative: abundant and
reactive ions like H$_3^+$ can react with H$_2$CO and CH$_3$OH to form
H$_2$COH$^+$ and CH$_3$OH$_2^+$, respectively. Upon dissociative
recombination with electrons, these species form HCO and CH$_3$O
together with other products. Estimated rate coefficients were used to
determine the [CH$_3$O]/[CH$_3$OH] and [HCO]/[H$_2$CO] abundance
ratios, together with experimental measurements of the products'
branching ratios for the dissociative recombination. We found that the
measured [CH$_3$O]/[CH$_3$OH] abundance ratio can be accounted
for well at steady state by this chemical route, and [HCO]/[H$_2$CO] is
also accounted well for by our model, provided that the branching
ratio for HCO in the dissociative recombination of H$_2$COH$^+$ is
close to 10\%. 
Our model also predicts abundances of H$_2$COH$^+$ and
CH$_3$OH$_2^+$, and measuring the abundances of these species
provides an important test of this reaction pathway. In the case of
CH$_3$OH$_2^+$, there is a need to first determine the frequencies of
the transitions in the millimetre domain. 

If the reactions we have proposed apply, the non-detection of
CH$_3$O in the protostar IRAS16293-2422 implies that the source is too
young for the ion-molecule to have significantly proceeded. In any
case, more observations of CH$_3$O in protostellar sources are
needed. 

Our model is based on several critical reactions, which involve key intermediate species, for which
there is an urgent need of experimental measurements, 
 the rates of which have not been
measured experimentally at low temperatures ($\sim 10\,$K), or for which branching ratios
are insufficiently constrained. These reactions are listed in Table\,\ref{reactionrates}.

Finally, there is a strong need to constrain the gas-phase
abundance of the isomer CH$_2$OH and to this effect, to determine its microwave spectrum, which is currently lacking.   
Our gas-phase scheme is consistent with  CH$_2$OH being less abundant than CH$_3$O. Should it be much more abundant, other processes would have to be at work than the gas-phase scheme we propose.

\begin{acknowledgements}
      This work has benefitted from the support of the CNRS programme "Physique et Chimie
du Milieu Interstellaire" (PCMI) . J.-C. Loison and K. Hickson are acknowledged for enlightening discussions about neutral-neutral reaction pathways. 
\end{acknowledgements}


\bibliographystyle{aa}
\bibliography{biblitex}

\begin{thebibliography}{96}
\expandafter\ifx\csname natexlab\endcsname\relax\def\natexlab#1{#1}\fi

\bibitem[{Aikawa {et~al.}(2003)Aikawa, Ohashi, \& Herbst}]{Aikawa:2003p1176}
Aikawa, Y., Ohashi, N., \& Herbst, E. 2003, ApJ, 593, 906

\bibitem[{{Andr{\'e}} {et~al.}(2014){Andr{\'e}}, {Di Francesco},
  {Ward-Thompson}, {Inutsuka}, {Pudritz}, \& {Pineda}}]{2014prpl.conf...27A}
{Andr{\'e}}, P., {Di Francesco}, J., {Ward-Thompson}, D., {et~al.} 2014,
  Protostars and Planets VI, 27

\bibitem[{Avery {et~al.}(1976)Avery, Broten, MacLeod, Oka, \&
  Kroto}]{Avery:1976eq}
Avery, L.~W., Broten, N.~W., MacLeod, J.~M., Oka, T., \& Kroto, H.~W. 1976,
  ApJ, 205, L173

\bibitem[{Avni(1976)}]{Avni:1976p3997}
Avni, Y. 1976, ApJ, 210, 642

\bibitem[{Bacmann {et~al.}(2010)Bacmann, Caux, Hily-Blant, Parise, Pagani,
  Bottinelli, Maret, Vastel, Ceccarelli, Cernicharo, Henning, Castets, Coutens,
  Bergin, Blake, Crimier, Demyk, Dominik, Gerin, Hennebelle, Kahane, Klotz,
  Melnick, Schilke, Wakelam, Walters, Baudry, Bell, Benedettini, Boogert,
  Cabrit, Caselli, Codella, Comito, Encrenaz, Falgarone, Fuente, Goldsmith,
  Helmich, Herbst, Jacq, Kama, Langer, Lefloch, Lis, Lord, Lorenzani, Neufeld,
  Nisini, Pacheco, Pearson, Phillips, Salez, Saraceno, Schuster, Tielens,
  van~der Tak, van~der Wiel, Viti, Wyrowski, Yorke, Faure, Benz, Coeur-Joly,
  Cros, G{\"u}sten, \& Ravera}]{Bacmann:2010p2163}
Bacmann, A., Caux, E., Hily-Blant, P., {et~al.} 2010, A{\&}A, 521, L42

\bibitem[{Bacmann {et~al.}(2002)Bacmann, Lefloch, Ceccarelli, Castets,
  Steinacker, \& Loinard}]{Bacmann:2002p1238}
Bacmann, A., Lefloch, B., Ceccarelli, C., {et~al.} 2002, A{\&}A, 389, L6

\bibitem[{Bacmann {et~al.}(2003)Bacmann, Lefloch, Ceccarelli, Steinacker,
  Castets, \& Loinard}]{Bacmann:2003p1218}
Bacmann, A., Lefloch, B., Ceccarelli, C., {et~al.} 2003, ApJ, 585, L55

\bibitem[{Bacmann {et~al.}(2012)Bacmann, Taquet, Faure, Kahane, \&
  Ceccarelli}]{Bacmann:2012p4015}
Bacmann, A., Taquet, V., Faure, A., Kahane, C., \& Ceccarelli, C. 2012, A{\&}A,
  541, L12

\bibitem[{{Balucani} {et~al.}(2015){Balucani}, {Ceccarelli}, \&
  {Taquet}}]{Balucani:2015p5368}
{Balucani}, N., {Ceccarelli}, C., \& {Taquet}, V. 2015, \mnras, 449, L16

\bibitem[{Banerji {et~al.}(2009)Banerji, Viti, Williams, \&
  Rawlings}]{Banerji:2009p4962}
Banerji, M., Viti, S., Williams, D.~A., \& Rawlings, J. M.~C. 2009, ApJ, 692,
  283

\bibitem[{Baulch {et~al.}(2005)Baulch, Bowman, Cobos, Cox, Just, Kerr, Pilling,
  Stocker, Troe, Tsang, Walker, \& Warnatz}]{Baulch:2005p5096}
Baulch, D.~L., Bowman, C.~T., Cobos, C.~J., {et~al.} 2005, J. Phys. Chem. Ref.
  Data, 34, 757

\bibitem[{Bizzocchi {et~al.}(2014)Bizzocchi, Caselli, Spezzano, \&
  Leonardo}]{Bizzocchi:2014p5365}
Bizzocchi, L., Caselli, P., Spezzano, S., \& Leonardo, E. 2014, A{\&}A, 569,
  A27

\bibitem[{Bottinelli {et~al.}(2004)Bottinelli, Ceccarelli, Neri, Williams,
  Caux, Cazaux, Lefloch, Maret, \& Tielens}]{Bottinelli:2004p1209}
Bottinelli, S., Ceccarelli, C., Neri, R., {et~al.} 2004, ApJ, 617, L69

\bibitem[{{Brady Ford} \& Shirley(2011)}]{BradyFord:2011p2416}
{Brady Ford}, A. \& Shirley, Y.~L. 2011, ApJ, 728, 144

\bibitem[{Brown {et~al.}(1988)Brown, Charnley, \& Millar}]{Brown:1988p3318}
Brown, P.~D., Charnley, S.~B., \& Millar, T.~J. 1988, MNRAS, 231, 409

\bibitem[{Brown {et~al.}(1975)Brown, Crofts, Godfrey, Gardner, Robinson, \&
  Whiteoak}]{Brown:1975p3219}
Brown, R.~D., Crofts, J.~G., Godfrey, P.~D., {et~al.} 1975, ApJ, 197, L29

\bibitem[{Caux {et~al.}(2011)Caux, Kahane, Castets, Coutens, Ceccarelli,
  Bacmann, Bisschop, Bottinelli, Comito, Helmich, Lefloch, Parise, Schilke,
  Tielens, van Dishoeck, Vastel, Wakelam, \& Walters}]{Caux:2011p2519}
Caux, E., Kahane, C., Castets, A., {et~al.} 2011, A{\&}A, 532, 23

\bibitem[{Cazaux {et~al.}(2003)Cazaux, Tielens, Ceccarelli, Castets, Wakelam,
  Caux, Parise, \& Teyssier}]{Cazaux:2003p2499}
Cazaux, S., Tielens, A. G. G.~M., Ceccarelli, C., {et~al.} 2003, ApJ, 593, L51

\bibitem[{{Cernicharo}(2012)}]{Cernicharo:2012p5353}
{Cernicharo}, J. 2012, in European Physical Journal Web of Conferences,
  Vol.~34, European Physical Journal Web of Conferences, 4002

\bibitem[{Cernicharo {et~al.}(2012)Cernicharo, Marcelino, Roueff, Gerin,
  Jim{\'e}nez-Escobar, \& Caro}]{Cernicharo:2012p4397}
Cernicharo, J., Marcelino, N., Roueff, E., {et~al.} 2012, ApJL, 759, L43

\bibitem[{Charnley {et~al.}(1992)Charnley, Tielens, \&
  Millar}]{Charnley:1992p3105}
Charnley, S.~B., Tielens, A. G. G.~M., \& Millar, T.~J. 1992, ApJ, 399, L71

\bibitem[{{Clary} {et~al.}(1985){Clary}, {Smith}, \& {Adams}}]{clary85}
{Clary}, D.~C., {Smith}, D., \& {Adams}, N.~G. 1985, Chem. Phys. Lett., 119,
  320

\bibitem[{Crapsi {et~al.}(2005)Crapsi, Caselli, Walmsley, Myers, Tafalla, Lee,
  \& Bourke}]{Crapsi:2005p30}
Crapsi, A., Caselli, P., Walmsley, C.~M., {et~al.} 2005, ApJ, 619, 379

\bibitem[{Crutcher(1979)}]{Crutcher:1979p4968}
Crutcher, R.~M. 1979, ApJ, 234, 881

\bibitem[{Dislaire {et~al.}(2012)Dislaire, Hily-Blant, Faure, Maret, Bacmann,
  \& {Pineau Des For{\^e}ts}}]{Dislaire:2012p4434}
Dislaire, V., Hily-Blant, P., Faure, A., {et~al.} 2012, A{\&}A, 537, 20

\bibitem[{Endo {et~al.}(1984)Endo, Saito, \& Hirota}]{Endo:1984p2512}
Endo, Y., Saito, S., \& Hirota, E. 1984, J. Chem. Phys., 81, 122

\bibitem[{Evans {et~al.}(2009)Evans, Dunham, J{\o}rgensen, Enoch, Mer{\'\i}n,
  van Dishoeck, Alcal{\'a}, Myers, Stapelfeldt, Huard, Allen, Harvey, Kempen,
  Blake, Koerner, Mundy, Padgett, \& Sargent}]{Evans:2009p4988}
Evans, N.~J., Dunham, M.~M., J{\o}rgensen, J.~K., {et~al.} 2009, ApJS, 181, 321

\bibitem[{Faure {et~al.}(2013)Faure, Hily-Blant, Gal, Rist, \& {Pineau Des
  For{\^e}ts}}]{Faure:2013p4435}
Faure, A., Hily-Blant, P., Gal, R.~L., Rist, C., \& {Pineau Des For{\^e}ts}, G.
  2013, ApJL, 770, L2

\bibitem[{Faure \& Lique(2012)}]{Faure:2012p5302}
Faure, A. \& Lique, F. 2012, MNRAS, 425, 740

\bibitem[{{Faure} {et~al.}(2010){Faure}, {Vuitton}, {Thissen}, {Wiesenfeld}, \&
  {Dutuit}}]{Faure:2010cy}
{Faure}, A., {Vuitton}, V., {Thissen}, R., {Wiesenfeld}, L., \& {Dutuit}, O.
  2010, Farad. Discuss., 147, 337

\bibitem[{Flower {et~al.}(2005)Flower, {Pineau Des For{\^e}ts}, \&
  Walmsley}]{Flower:2005p1891}
Flower, D.~R., {Pineau Des For{\^e}ts}, G., \& Walmsley, C.~M. 2005, A{\&}A,
  436, 933

\bibitem[{Flower {et~al.}(2006)Flower, {Pineau Des For{\^e}ts}, \&
  Walmsley}]{Flower:2006p1295}
Flower, D.~R., {Pineau Des For{\^e}ts}, G., \& Walmsley, C.~M. 2006, A{\&}A,
  449, 621

\bibitem[{Frau {et~al.}(2012)Frau, Girart, \& Beltr{\'a}n}]{Frau:2012p3985}
Frau, P., Girart, J.~M., \& Beltr{\'a}n, M.~T. 2012, A{\&}A, 537, L9

\bibitem[{Friberg {et~al.}(1988)Friberg, Hjalmarson, Madden, \&
  Irvine}]{Friberg:1988p3131}
Friberg, P., Hjalmarson, A., Madden, S.~C., \& Irvine, W.~M. 1988, A{\&}A, 195,
  281

\bibitem[{{Garrod} {et~al.}(2006){Garrod}, {Park}, {Caselli}, \&
  {Herbst}}]{Garrod:2006p4851}
{Garrod}, R., {Park}, I.~H., {Caselli}, P., \& {Herbst}, E. 2006, Farad.
  Discuss., 133, 51

\bibitem[{Garrod \& Herbst(2006)}]{Garrod:2006p2498}
Garrod, R.~T. \& Herbst, E. 2006, A{\&}A, 457, 927

\bibitem[{Garrod {et~al.}(2007)Garrod, Wakelam, \& Herbst}]{Garrod:2007p2821}
Garrod, R.~T., Wakelam, V., \& Herbst, E. 2007, A{\&}A, 467, 1103

\bibitem[{Garrod {et~al.}(2008)Garrod, Weaver, \& Herbst}]{Garrod:2008p2520}
Garrod, R.~T., Weaver, S. L.~W., \& Herbst, E. 2008, ApJ, 682, 283

\bibitem[{{Geppert} {et~al.}(2006){Geppert}, {Hamberg}, {Thomas},
  {{\"O}sterdahl}, {Hellberg}, {Zhaunerchyk}, {Ehlerding}, {Millar}, {Roberts},
  {Semaniak}, {Ugglas}, {K{\"a}llberg}, {Simonsson}, {Kaminska}, \&
  {Larsson}}]{Geppert:2006p4196}
{Geppert}, W.~D., {Hamberg}, M., {Thomas}, R.~D., {et~al.} 2006, Farad.
  Discuss., 133, 177

\bibitem[{Gerin {et~al.}(2009)Gerin, Goicoechea, Pety, \&
  Hily-Blant}]{Gerin:2009p4072}
Gerin, M., Goicoechea, J.~R., Pety, J., \& Hily-Blant, P. 2009, A{\&}A, 494,
  977

\bibitem[{Goldsmith \& Li(2005)}]{Goldsmith:2005p5113}
Goldsmith, P.~F. \& Li, D. 2005, ApJ, 622, 938

\bibitem[{{G{\'o}mez Mart{\'\i}n} {et~al.}(2014){G{\'o}mez Mart{\'\i}n},
  Caravan, Blitz, Heard, \& Plane}]{gomezmartin:2014p4918}
{G{\'o}mez Mart{\'\i}n}, J., Caravan, R., Blitz, M., Heard, D., \& Plane, J.
  2014, J. Phys. Chem. A, 118, 2693

\bibitem[{{Gottlieb}(1973)}]{Gottlieb:1973p5025}
{Gottlieb}, C.~A. 1973, in Molecules in the Galactic Environment, ed. M.~A.
  {Gordon} \& L.~E. {Snyder}, 181

\bibitem[{Gutman {et~al.}(1982)Gutman, Sanders, \& Butler}]{Gutman:1982gv}
Gutman, D., Sanders, N., \& Butler, J.~E. 1982, J. Phys. Chem., 86, 66

\bibitem[{Guzm{\'a}n {et~al.}(2013)Guzm{\'a}n, Goicoechea, Pety, Gratier,
  Gerin, Roueff, Petit, Bourlot, \& Faure}]{Guzman:2013p4399}
Guzm{\'a}n, V.~V., Goicoechea, J.~R., Pety, J., {et~al.} 2013, A{\&}A, 560, 73

\bibitem[{Hamberg {et~al.}(2007)Hamberg, Geppert, Thomas, Zhaunerchyk,
  {\"O}sterdahl, Ehlerding, Kaminska, Semaniak, Ugglas, K{\"a}llberg, Paal,
  Simonsson, \& Larsson}]{Hamberg:2007p4982}
Hamberg, M., Geppert, W.~D., Thomas, R.~D., {et~al.} 2007, Molecular Physics,
  105, 899

\bibitem[{Hasegawa \& Herbst(1993)}]{Hasegawa:1993tv}
Hasegawa, T.~I. \& Herbst, E. 1993, MNRAS, 263, 589

\bibitem[{Hily-Blant {et~al.}(2013)Hily-Blant, {Pineau Des For{\^e}ts}, Faure,
  Gal, \& Padovani}]{HilyBlant:2013p4440}
Hily-Blant, P., {Pineau Des For{\^e}ts}, G., Faure, A., Gal, R.~L., \&
  Padovani, M. 2013, A{\&}A, 557, 65

\bibitem[{Horn {et~al.}(2004)Horn, M{\o}llendal, Sekiguchi, Uggerud, Roberts,
  Herbst, Viggiano, \& Fridgen}]{Horn:2004p2926}
Horn, A., M{\o}llendal, H., Sekiguchi, O., {et~al.} 2004, ApJ, 611, 605

\bibitem[{Li \& Goldsmith(2003)}]{Li:2003jz}
Li, D. \& Goldsmith, P.~F. 2003, ApJ, 585, 823

\bibitem[{Loinard {et~al.}(2000)Loinard, Castets, Ceccarelli, Tielens, Faure,
  Caux, \& Duvert}]{Loinard:2000p4986}
Loinard, L., Castets, A., Ceccarelli, C., {et~al.} 2000, A{\&}A, 359, 1169

\bibitem[{Loison {et~al.}(2014)Loison, Wakelam, Hickson, Bergeat, \&
  Mereau}]{Loison:2014p4321}
Loison, J.-C., Wakelam, V., Hickson, K.~M., Bergeat, A., \& Mereau, R. 2014,
  MNRAS, 437, 930

\bibitem[{McElroy {et~al.}(2013)McElroy, Walsh, Markwick, Cordiner, Smith, \&
  Millar}]{McElroyD:2013ki}
McElroy, D., Walsh, C., Markwick, A.~J., {et~al.} 2013, A{\&}A, 550, A36

\bibitem[{M{\"u}ller {et~al.}(2004)M{\"u}ller, Menten, \&
  M{\"a}der}]{Muller:2004p4299}
M{\"u}ller, H. S.~P., Menten, K.~M., \& M{\"a}der, H. 2004, A{\&}A, 428, 1019

\bibitem[{{M{\"u}ller} {et~al.}(2005){M{\"u}ller}, {Schl{\"o}der}, {Stutzki},
  \& {Winnewisser}}]{Muller:2005p4298}
{M{\"u}ller}, H.~S.~P., {Schl{\"o}der}, F., {Stutzki}, J., \& {Winnewisser}, G.
  2005, J. Mol. Struct., 742, 215

\bibitem[{M{\"u}ller {et~al.}(2001)M{\"u}ller, Thorwirth, Roth, \&
  Winnewisser}]{Muller:2001p2418}
M{\"u}ller, H. S.~P., Thorwirth, S., Roth, D.~A., \& Winnewisser, G. 2001,
  A{\&}A, 370, L49

\bibitem[{Ohishi {et~al.}(1996)Ohishi, Ishikawa, Amano, Oka, Irvine, Dickens,
  Ziurys, \& Apponi}]{Ohishi:1996p4278}
Ohishi, M., Ishikawa, S.-I., Amano, T., {et~al.} 1996, ApJL v.471, 471, L61

\bibitem[{Osamura {et~al.}(2005)Osamura, Roberts, \&
  Herbst}]{Osamura:2005p4965}
Osamura, Y., Roberts, H., \& Herbst, E. 2005, ApJ, 621, 348

\bibitem[{Padovani {et~al.}(2011)Padovani, Walmsley, Tafalla, Hily-Blant, \&
  {Pineau Des For{\^e}ts}}]{Padovani:2011p4178}
Padovani, M., Walmsley, C.~M., Tafalla, M., Hily-Blant, P., \& {Pineau Des
  For{\^e}ts}, G. 2011, A{\&}A, 534, 77

\bibitem[{Pagani {et~al.}(2009)Pagani, Vastel, Hugo, Kokoouline, Greene,
  Bacmann, Bayet, Ceccarelli, Peng, \& Schlemmer}]{Pagani:2009p1030}
Pagani, L., Vastel, C., Hugo, E., {et~al.} 2009, A{\&}A, 494, 623

\bibitem[{Palmer {et~al.}(1969)Palmer, Zuckerman, Buhl, \&
  Snyder}]{Palmer:1969p4000}
Palmer, P., Zuckerman, B., Buhl, D., \& Snyder, L.~E. 1969, ApJ, 156, L147

\bibitem[{Parise {et~al.}(2004)Parise, Castets, Herbst, Caux, Ceccarelli,
  Mukhopadhyay, \& Tielens}]{Parise:2004p1297}
Parise, B., Castets, A., Herbst, E., {et~al.} 2004, A{\&}A, 416, 159

\bibitem[{Peters {et~al.}(2013)Peters, Duflot, Wiesenfeld, \&
  Toubin}]{Peters:2013p4920}
Peters, P.~S., Duflot, D., Wiesenfeld, L., \& Toubin, C. 2013, J. Chem. Phys.,
  139, 4310

\bibitem[{Pickett {et~al.}(1998)Pickett, Poynter, Cohen, Delitsky, Pearson, \&
  M{\"u}ller}]{Pickett:1998p4433}
Pickett, H.~M., Poynter, R.~L., Cohen, E.~A., {et~al.} 1998, JQSRT, 60, 883

\bibitem[{Pirim \& Krim(2011{\natexlab{a}})}]{Pirim:2011p4136}
Pirim, C. \& Krim, L. 2011{\natexlab{a}}, Chem. Phys., 380, 67

\bibitem[{Pirim \& Krim(2011{\natexlab{b}})}]{Pirim:2011p4905}
Pirim, C. \& Krim, L. 2011{\natexlab{b}}, Phys. Chem. Chem. Phys., 13, 19454

\bibitem[{Pirim {et~al.}(2010)Pirim, Krim, Laffon, Parent, Pauzat, Pilm{\'e},
  \& Ellinger}]{Pirim:2010p4906}
Pirim, C., Krim, L., Laffon, C., {et~al.} 2010, J. Phys. Chem. A, 114, 3320

\bibitem[{Rabli \& Flower(2010)}]{Rabli:2010p4127}
Rabli, D. \& Flower, D.~R. 2010, MNRAS, 406, 95

\bibitem[{Roy {et~al.}(2014)Roy, Andr{\'e}, Palmeirim, Attard, K{\"o}nyves,
  Schneider, Peretto, Men'shchikov, Ward-Thompson, Kirk, Griffin, Marsh,
  Abergel, Arzoumanian, Benedettini, Hill, Motte, Luong, Pezzuto,
  Rivera-Ingraham, Roussel, Rygl, Spinoglio, Stamatellos, \&
  White}]{Roy:2014p4845}
Roy, A., Andr{\'e}, P., Palmeirim, P., {et~al.} 2014, A{\&}A, 562, 138

\bibitem[{Ruaud {et~al.}(2015)Ruaud, Loison, Hickson, Gratier, Hersant, \&
  Wakelam}]{Ruaud:2015p5367}
Ruaud, M., Loison, J.~C., Hickson, K.~M., {et~al.} 2015, MNRAS, 447, 4004

\bibitem[{Saito(1972)}]{Saito:1972p4219}
Saito, S. 1972, ApJ, 178, L95

\bibitem[{Sayah {et~al.}(1988)Sayah, Li, Caballero, \&
  Jackson}]{Sayah:1988p4966}
Sayah, N., Li, X., Caballero, J., \& Jackson, W. 1988, J. Photochem. Photobiol.
  A, 45, 177

\bibitem[{Schenewerk {et~al.}(1988)Schenewerk, Jewell, Snyder, Hollis, \&
  Ziurys}]{Schenewerk:1988p4307}
Schenewerk, M.~S., Jewell, P.~R., Snyder, L.~E., Hollis, J.~M., \& Ziurys,
  L.~M. 1988, ApJ, 328, 785

\bibitem[{Sch{\"o}ier {et~al.}(2002)Sch{\"o}ier, J{\o}rgensen, Dishoeck, \&
  Blake}]{Schoier:2002p4241}
Sch{\"o}ier, F.~L., J{\o}rgensen, J.~K., Dishoeck, E. F.~V., \& Blake, G.~A.
  2002, A{\&}A, 390, 1001

\bibitem[{Shannon {et~al.}(2013)Shannon, Blitz, Goddard, \&
  Heard}]{Shannon:2013p4217}
Shannon, R.~J., Blitz, M.~A., Goddard, A., \& Heard, D.~E. 2013, Nature Chem,
  5, 745

\bibitem[{{Shannon} {et~al.}(2014){Shannon}, {Cossou}, {Loison}, {Caubet},
  {Balucani}, {Seakins}, {Wakelam}, \& {Hickson}}]{Shannon:2014aa}
{Shannon}, R.~J., {Cossou}, C., {Loison}, J.-C., {et~al.} 2014, RSC Adv., 4,
  26342

\bibitem[{Sims {et~al.}(1994)Sims, Queffelec, Defrance, Rebrion-Rowe, Travers,
  Bocherel, Rowe, \& Smith}]{Sims:1994p5303}
Sims, I.~R., Queffelec, J.-L., Defrance, A., {et~al.} 1994, J. Chem. Phys.,
  100, 4229

\bibitem[{Snyder {et~al.}(1976)Snyder, Hollis, \& Ulich}]{Snyder:1976p4306}
Snyder, L.~E., Hollis, J.~M., \& Ulich, B.~L. 1976, ApJ, 208, L91

\bibitem[{Tafalla {et~al.}(2004)Tafalla, Myers, Caselli, \&
  Walmsley}]{Tafalla:2004p1027}
Tafalla, M., Myers, P.~C., Caselli, P., \& Walmsley, C.~M. 2004, A{\&}A, 416,
  191

\bibitem[{Tafalla \& Santiago(2004)}]{Tafalla:2004p3983}
Tafalla, M. \& Santiago, J. 2004, A{\&}A, 414, L53

\bibitem[{Tafalla {et~al.}(2006)Tafalla, Santiago-Garc{\'\i}a, Myers, Caselli,
  Walmsley, \& Crapsi}]{Tafalla:2006p3988}
Tafalla, M., Santiago-Garc{\'\i}a, J., Myers, P.~C., {et~al.} 2006, A{\&}A,
  455, 577

\bibitem[{Taquet {et~al.}(2012)Taquet, Ceccarelli, \&
  Kahane}]{Taquet:2012p5062}
Taquet, V., Ceccarelli, C., \& Kahane, C. 2012, A{\&}A, 538, 42

\bibitem[{{Tielens}(2010)}]{Tielens:2005p4957}
{Tielens}, A.~G.~G.~M. 2010, {The Physics and Chemistry of the Interstellar
  Medium} (UK: Cambridge University Press)

\bibitem[{{Tielens} \& {Hagen}(1982)}]{Tielens:1982p3002}
{Tielens}, A.~G.~G.~M. \& {Hagen}, W. 1982, \aap, 114, 245

\bibitem[{Townes \& Cheung(1969)}]{Townes:1969p5328}
Townes, C.~H. \& Cheung, A.~C. 1969, ApJ, 157, L103

\bibitem[{Troland \& Crutcher(2008)}]{Troland:2008p1151}
Troland, T.~H. \& Crutcher, R.~M. 2008, ApJ, 680, 457

\bibitem[{Tucker {et~al.}(1974)Tucker, Kutner, \& Thaddeus}]{Tucker:1974fs}
Tucker, K.~D., Kutner, M.~L., \& Thaddeus, P. 1974, ApJ, 193, L115

\bibitem[{van~der Tak {et~al.}(2007)van~der Tak, Black, Sch{\"o}ier, Jansen, \&
  Dishoeck}]{vanderTak:2007p4401}
van~der Tak, F. F.~S., Black, J.~H., Sch{\"o}ier, F.~L., Jansen, D.~J., \&
  Dishoeck, E. F.~V. 2007, A{\&}A, 468, 627

\bibitem[{van Dishoeck {et~al.}(1995)van Dishoeck, Blake, Jansen, \&
  Groesbeck}]{vanDishoeck:1995p3020}
van Dishoeck, E.~F., Blake, G.~A., Jansen, D.~J., \& Groesbeck, T.~D. 1995,
  ApJ, 447, 760

\bibitem[{Vastel {et~al.}(2014)Vastel, Ceccarelli, Lefloch, \&
  Bachiller}]{Vastel:2014p5375}
Vastel, C., Ceccarelli, C., Lefloch, B., \& Bachiller, R. 2014, ApJL, 795, L2

\bibitem[{Vasyunin \& Herbst(2013)}]{Vasyunin:2013p4398}
Vasyunin, A.~I. \& Herbst, E. 2013, ApJ, 769, 34

\bibitem[{Wakelam {et~al.}(2012)Wakelam, Herbst, Loison, Smith, Chandrasekaran,
  Pavone, Adams, Bacchus-Montabonel, Bergeat, B{\'e}roff, Bierbaum, Chabot,
  Dalgarno, Dishoeck, Faure, Geppert, Gerlich, Galli, H{\'e}brard, Hersant,
  Hickson, Honvault, Klippenstein, Picard, Nyman, Pernot, Schlemmer, Selsis,
  Sims, Talbi, Tennyson, Troe, Wester, \& Wiesenfeld}]{Wakelam:2012p4970}
Wakelam, V., Herbst, E., Loison, J.-C., {et~al.} 2012, ApJS, 199, 21

\bibitem[{Wang {et~al.}(1973)Wang, Eyre, \& Dorfman}]{Wang:1973p4921}
Wang, H.~Y., Eyre, J.~A., \& Dorfman, L.~M. 1973, J. Chem. Phys., 59, 5199

\bibitem[{Wang \& Bowie(2012)}]{Wang:2012dja}
Wang, T. \& Bowie, J.~H. 2012, Org. Biomol. Chem., 10, 3219

\bibitem[{Watanabe \& Kouchi(2002)}]{Watanabe:2002p3383}
Watanabe, N. \& Kouchi, A. 2002, ApJ, 571, L173

\bibitem[{Xu \& Lovas(1997)}]{Xu:1997p4300}
Xu, L.-H. \& Lovas, F.~J. 1997, J. Phys. Chem. Ref. Data, 26, 17

\end{thebibliography}
\Online

\begin{appendix}

\section{The ``locked dipole `` approximation}\label{capturerates}

Our simple steady-state chemical model (detailed in sections~4.2 and
4.3) is subtantially based on the loss of HCO, H$_2$CO, CH$_3$O and
CH$_3$OH by reactions with H$_3^+$. To the best of our knowledge, the
corresponding rate coefficients have not been measured
experimentally. It should be noted that the dipole moments of these
species are quite large: 1.53~D for HCO, 2.33~D for H$_2$CO, 2.12~D
for CH$_3$O and 1.67~D for CH$_3$OH (JPL catalog). The rates are
therefore expected to significantly exceed the Langevin rate of $\sim
10^{-9}$~cm$^3$s$^{-1}$. Several theoretical studies have shown,
however, that the contribution of the ion–dipole interaction is
difficult to estimate \citep[see][and references therein]{Faure:2010cy}. In the case of the similar HCN+H$_3^+$ reaction, the popular
inifinite-order sudden (IOS) and average-dipole-orientation (ADO)
theories were found to underestimate the rate coefficient at low
temperature by large factors \citep{clary85}. For such polar systems withrelatively large rotational constants ($B\gtrsim 1$~cm$^{-1}$), the
``locked dipole'' approximation provides better results in the low
temperature regime because the short rotational period allows dipole
alignment. Indeed, for a typical rate coefficients of $5\times
10^{-8}$~cm$^3$s$^{-1}$, one can estimate an interaction time of
$t_{int}\sim 30$~ps at 10~K, which is similar or larger than the
typical rotational periods. The locked-dipole formula can be found
e.g. in Eq.~(1) of \citet{Faure:2010cy}. The corresponding values for
the rate coefficients at 10~K are given in Table\,\ref{reactionrates}.

\end{appendix}


\end{document}